\documentclass{aa}  

\usepackage{subcaption}
\usepackage{graphicx}
\usepackage{lastpage}
\usepackage{xcolor}
\usepackage{natbib}
\usepackage{txfonts}
\usepackage{rotating}
\usepackage{hyperref}
\hypersetup{colorlinks=true,citecolor=blue,linkcolor=blue}

\newcommand{\HI}{H\,{\sc i}}

\begin{document}

   \title{MUSE-ALMA Haloes} 
   \subtitle{XVI. A census of the condensed baryons in $z\sim0.5$ \HI\ absorption-selected galaxies}

   \author{T. O'Beirne
          \inst{1,2,3}\fnmsep\thanks{email: tamsynobeirne.astro@gmail.com}
          \and
          C.~Péroux\inst{1,4}
          \and
          M.~Zwaan\inst{1}
          \and
          C.~Barfety\inst{5}
          \and 
          V.~Bollo\inst{6}
          \and
          J.~Chen\inst{5}
          \and
          A. L. Danhaive\inst{7,8}
          \and
          N.~M.~F\"orster~Schreiber\inst{5}
          \and
          L.~Tacconi\inst{5}
          \and
          S.~Weng\inst{4}
          \and
          V.~Kulkarni\inst{9}
          \and
          L. Chemin\inst{10}
          \and
          R. Augustin\inst{11}
          \and
          B. Casavecchia\inst{12}
          \and
          M.~E.~Cluver\inst{2} 
          \and
          M.~Halley\inst{9}
          \and
          A. Karki\inst{9}
          \and
          V.~A.~Kilborn\inst{2}
          \and
          O.~I.~Wong\inst{3,13}
          }

   \institute{European Southern Observatory, Karl-Schwarzschild-Str. 2, 85748 Garching near Munich, Germany  %1    
         \and
             Centre for Astrophysics and Supercomputing, Swinburne University of Technology, Hawthorn, Victoria 3122, Australia %2
        \and
        CSIRO Space \& Astronomy, PO Box 1130, Bentley, Western Australia 6102, Australia %3
        \and
        Aix Marseille Univ., CNRS, LAM, (Laboratoire d’Astrophysique de Marseille), UMR 7326, F-13388 Marseille, France %4
        \and
        Max-Planck-Institut für extraterrestrische Physik, Gießenbachstraße, Garching, 85748, Germany %5  
        \and
        Department of Astronomy, Stockholm University, Oscar Klein Centre, AlbaNova University Centre, 106 91 Stockholm, Sweden % 6
        \and
        Kavli Institute for Cosmology, University of Cambridge, Madingley Road, Cambridge CB3 0HA, UK %7
        \and
        Cavendish Laboratory, University of Cambridge, 19 JJ Thomson Avenue, Cambridge CB3 0HE, UK % 8
        \and
        Department of Physics and Astronomy, University of South Carolina, Columbia, SC 29208, USA %9
        \and
        Observatoire Astronomique de Strasbourg - UMR 7550, 11 rue de l’Université, 67000 Strasbourg, France %10
        \and
        Leibniz-Institut für Astrophysik Potsdam (AIP), An der Sternwarte 16, 14482 Potsdam, Germany %11
        \and
        Max-Planck-Institut für Astrophysik, Karl-Schwarzschild-Strasse 1, 85748 Garching near Munich, Germany %12
        \and 
        International Centre for Radio Astronomy Research, The University of Western Australia M468, 35 Stirling Highway, Crawley WA 6009, Australia %13
             }

   \date{Received XXX; accepted XXX}

% \abstract{}{}{}{}{} 
% 5 {} token are mandatory
 
  \abstract{
We use the MUSE-ALMA Haloes sample to build a census of the condensed baryons (stars and cold gas) to study the interaction of different gas phases in the interstellar medium and circumgalactic medium (CGM). The sample includes 79 gas-rich galaxies at $0.2<z<1.2$ located near strong neutral atomic hydrogen (\HI) Ly-$\alpha$ absorbers ($N_{\rm HI}>10^{18}$ cm$^{-2}$) along quasar sightlines within projected separations of $\sim5-270$ kpc and radial velocity separations of up to 500 km s$^{-1}$.  HST and ground-based spectroscopy characterise the absorbers, while HST imaging, MUSE and ALMA observations trace the stellar populations, ionised gas, and molecular gas. We find the following key results. (i) \HI\ masses inferred from optical galaxy properties show broad agreement with those derived from the CGM absorber properties for impact parameters $<100$ kpc, although group environments can significantly influence these estimates. (ii)  The molecular gas fraction declines more steeply with stellar mass than the \HI\ fraction. This suggests that stellar mass builds up more rapidly than the cold gas reservoir, while massive galaxies either consume their molecular gas more efficiently or convert \HI\ into H$_2$ less efficiently. (iii) Galaxies with high stellar surface densities exhibit the highest baryon fractions. This may reflect baryonic feedback, which can produce flatter inner dark matter profiles in high surface density systems, or aperture effects, where the ionised gas kinematic tracer probes only the innermost, baryon-dominated regions in compact sources, but extends further into the dark matter halo in sources with extended baryonic distributions. (iv) A few galaxies are consistent with being the baryon-dominated descendants of the low mass $z\sim6$ galaxies detected by JWST, while many appear to be substantially more dark matter dominated, lower mass systems with progenitors beyond JWST detection limits. This study provides a multiscale, multiphase view of the gas in and around galaxies, allowing us to probe the baryon budget across cosmic time.
}


   \maketitle
%
%-------------------------------------------------------------------

%%%%%%%%%%%%%%%%%%%%%%%%%%%%%%%%%%%%%%%%%%%%%%%%%%%%%%%
\section{Introduction}

  A complete census of the baryons in and around galaxies is essential for understanding galaxy evolution, as it directly traces the processes governing the galactic baryon cycle. This cycle describes the interplay between gas availability and star formation, and is regulated by processes such as gas accretion, feedback from star formation and active galactic nuclei (AGN),  gas outflows and interactions with the environment \citep[e.g.][]{Dave:11,Lilly:13,Peroux:20,Tacconi:20,Cortese:21}. The diffuse, multiphase gas that surrounds the galaxy out to the virial radius is called the circumgalactic medium (CGM), and this serves as the primary site for gas exchange between the interstellar medium (ISM) and the intergalactic medium (IGM) \citep[][]{Borthakur:15,Tumlinson:17,Peroux:24}. 

 {Multi-wavelength observations are essential, as each tracer reveals distinct structures, gas phases  {and kinematics} that together map the galactic baryon cycle.} Molecular gas is typically concentrated in the central regions of galaxies and acts as the direct fuel for star formation.   Neutral atomic hydrogen (\HI) dominates the cold gas reservoir of a galaxy. It often extends well beyond the optical disc, and has been shown to extend into the CGM as diffuse gas \citep[e.g.][]{Borthakur:16,Ramesh:24,Das:24}.  Supernovae and stellar winds inject energy into the ISM, heating, ionising, and enriching the gas with metals. These processes, along with feedback from AGN, can drive ionised gas outflows. Such outflows may enrich the CGM with metals or recycle material back into the galaxy via the galactic fountain effect \citep{Veilleux:05,Fraternali:17, Tumlinson:17,Mitchell:20}. The gas in and around galaxies is thus multiphase, consisting of gas at a range of temperatures, ionisation states and densities.

 {Even with extensive multi-wavelength observations, assembling a complete inventory of baryons across all gas phases and environments remains challenging. The discrepancy is known as the `missing baryon problem', which has two components: (1) the global census of baryons in galaxies, clusters, and the intergalactic medium falls short of the cosmic density, and (2) individual galaxy haloes retain fewer baryons than expected from the cosmic density \citep{Persic:92,McGaugh:08,McGaugh:10,Werk:14}.  {Studies have shown that $\sim30\%$ of baryons are unaccounted for in global surveys \citep{Shull:12,degraaff:19}, while previous estimates suggested that} $\sim60\%$ were missing from Milky Way mass ($\sim10^{12}$~M$_\odot$) haloes \citep{Werk:14}. Condensed baryons in galaxy discs, (i.e. molecular gas, atomic hydrogen, dust and stars) represent only a small fraction of the total, whereas warmer ($\gtrsim10^4$~K), diffuse gas is expected to dominate the baryon budget. Much of this warm and hot gas is thought to reside in the CGM. %\citep{Bregman:07,Werk:14,Zhang:26}.  
 
 {Absorption-line studies in the spectra of bright background quasars can probe the diffuse component. \cite{Werk:14} show that $\sim 10^4$ K CGM gas accounts for $25\%-45\%$ ($>6.5\times10^{10}$ M$_{\odot}$) of the total baryon content of Milky Way-mass Haloes. The warm and hot CGM ($\gtrsim 10^4$ K) are more poorly constrained, however, they are expected to host much of the remaining baryons. Using stacked eROSITA observations, \cite{Zhang:26} construct the average X-ray surface brightness profile of the CGM around Milky Way-mass galaxies. They infer that the hot CGM contributes $0.8-3.5\times10^{11}$ M$_\odot$ to the baryon budget.}

 {This multiphase gas in and around galaxies does not exist in isolation, but is shaped by its surroundings.} The local environment can enhance or disrupt the baryon cycle driven by internal processes within galaxies. In fact, most galaxies in the local Universe reside in groups \citep[$\sim40-55\%$;][]{Eke:04,Robotham:11}. Environmental processes can be broadly divided into two categories: galaxy-medium interactions, such as ram pressure stripping \citep[RPS;][]{Gunn:72}, and galaxy-galaxy interactions, including tidal stripping \citep[e.g.][]{Kilborn:06}. Consequently, galaxy groups often contain substantial reservoirs of intragroup \HI\ gas \citep[e.g.][]{obeirne:24}. While local interactions can trigger enhanced star formation \citep[e.g.][]{Patton:13}, more massive groups, with increasing halo mass, tend to promote star formation quenching \citep{Cluver:20,VanKempen:24}.

Selecting galaxies in \HI\ offers a unique view of the Universe, favouring gas-rich late-type systems in  {low-density field and group environments over dense clusters.} Unlike optical selection, an \HI-selected sample detects galaxies independent of their stellar mass or optical surface brightness, revealing populations that are otherwise missed. At low redshifts ($z\lesssim0.4$), \HI\ can be observed directly through its 21 cm emission line, and \HI-bearing galaxies are well characterised by large untargeted surveys such as the Widefield ASKAP L-band Legacy All-sky Blind surveY \citep[WALLABY;][]{Koribalski:20}. Beyond redshift $z\sim0.1$, the 21 cm emission from individual galaxies becomes increasingly difficult to detect within reasonable integration times. This restricts observations to small regions with deep integrations, such as surveys like Looking at the Distant Universe with MeerKAT Array \citep[LADUMA;][]{Blyth:16}, or to statistical approaches such as stacking experiments \citep[e.g.][]{Sinigaglia:22,Chowdhury:24}. 

An alternative and effective way to study the cool ($\lesssim10^4$~K) gas in galaxies at higher redshifts is through \HI\ Ly$\alpha$ or low-ionisation metal-line absorption against bright background sources. This method is sensitive to diffuse gas, and its sensitivity is independent of the redshift of the intervening absorber. Such approaches are commonly applied at $z\lesssim0.4$  \citep[e.g.][]{Tripp:98,Prochaska:11,Werk:13}, and at $z\sim2-3$ \citep[e.g.][]{Steidel:10,Rudie:12,Turner:14}. However, there remains a gap between these two epochs that must be addressed in order to build a complete picture of the evolution of neutral gas in galaxies. This gap has begun to be explored in recent studies, including the MUSE-ALMA Haloes programme \citep{Peroux:22}, {the MUSE Ultra Deep Field survey \citep[MUDF;][]{Fossati:19}, } the MUSE Quasar-field Blind Emitters Survey \citep[MUSEQuBES;][]{Muzahid:20}, the Cosmic Ultraviolet Baryon Survey \citep[CUBS;][]{Chen:20}, the Bimodal Absorption System Imaging Campaign \citep[BASIC;][]{Berg:23}, and the MusE GAs FLOw and Wind survey \citep[MEGAFLOW;][]{Bouche:25}. 

 {The MUSE–ALMA Haloes programme provides a unique opportunity to bridge this redshift gap, with its strength lying in its multiwavelength coverage.} The programme is comprised of 79 galaxies located within 0.5 arcmin and 500 km s$^{-1}$ of 32 \HI\ Ly$\alpha$ absorbers ($N_{\rm HI}>10^{18}$ cm$^{-2}$) across a redshift range of $0.2<z<1.2$.   {HST and ground-based spectroscopy characterise the absorbers, while HST imaging, MUSE and ALMA observations trace the stellar populations, ionised gas and molecular gas. Crucially, the programme enables a direct connection between absorber properties and the physical characteristics of their nearby galaxies.}

This paper brings together the  {extensive and complementary} data collected on this galaxy sample over the past several years to characterise the baryonic content and gas phases. We aim to investigate how the baryon cycle operates in \HI-selected galaxy overdensities at $0.2<z<1.2$, and examine how the molecular, atomic, and ionised gas phases interact across the CGM and ISM to regulate star formation. The paper is structured as follows. Section \ref{sec:data} summarises the multiwavelength observations. In Sect. \ref{sec:quantities}, we outline the methods used to infer the galaxy properties.  {Section \ref{sec:baryon_res} presents the ISM and CGM as the primary baryon reservoirs, and Sect. \ref{sec:fractions} examines} the condensed baryon budget of the MUSE-ALMA Haloes galaxies, including the gas and baryon fractions, and places them in context of other galaxy samples across redshift.  Finally, Sect. \ref{sec:concl} summarises our findings and presents our conclusions. Throughout the paper we assume a cosmology with H$_0$ = 70 km s$^{-1}$ Mpc$^{-1}$ and $\Omega_{m,0} = 0.3$. We adopt the AB magnitude system and use velocity in the optical convention.

%%%%%%%%%%%%%%%%%%%%%%%%%%%%%%%%%%%%%%%%%%%%%%%%%%%%%%%
\section{Data}
\label{sec:data}

The source selection, multiwavelength observations, and data reduction are described in detail by \cite{Peroux:22}, \cite{Weng:23a}, \cite{Bollo:26} and \cite{Peroux:26}. In this section, we summarise the key aspects of the MUSE–ALMA Haloes dataset relevant to the following baryon census.

\subsection{HST}
{To construct the MUSE-ALMA Haloes sample, HST UV spectroscopy was used to identify 32 \HI\ Ly$\alpha$ absorbers in 19 background quasar sightlines. The HST spectra were obtained from a combination of archival data with spectral resolutions of  $R= 20,000-30,000$ \citep[][ and references therin]{Peroux:22}.  }

In addition, most fields are covered by multi-band imaging, providing high spatial resolution  {($\sim 0.07-0.09$ arcsec)}.  {These data come} from a 40-orbit medium programme (ID: 15939; PI: C. Peroux) with the Wide Field Camera 3, with optical (UVIS) and infrared (IR) detectors, as well as archival data. {Eighteen of the 19 fields have coverage in at least one broadband filter, and 16 of these} have at least three broad-band filters available, with the specific choice of filters optimised for the redshift of the absorber. { In fields with several absorbers, the chosen broad-band filters are well suited to probe the wavelength range bluewards of the 4000$\AA$ break. All 19 fields are included in the MUSE-ALMA Haloes programme; however, fields lacking coverage in three HST bands are supplemented with Dark Energy Camera Legacy Survey \citep[DECaLS;][]{Dey:19} data to enable spectral energy distribution (SED) fitting (see Sect. \ref{sec:lit_prop}). }

\subsection{Additional ground-based quasar spectroscopy}

Thirty-one of the 32 \HI\ absorbers have additional high resolution ground-based spectroscopy from VLT/UVES (PI: Péroux, ID: 113.A-0369; {$R=40,000-110,000$}), Keck/HIRES {($R=25,000-85,000$)} or VLT/X-Shooter {($R=4000-17,000$)} archival data  {(Halley et al., in prep.)}. This provides metal absorption lines such as Fe {\sc ii}, Si {\sc ii}, Zn {\sc ii} and C {\sc iv}. {Absorber redshifts are determined from metal lines where available, and otherwise from HST Ly$\alpha$ absorption. These precise redshifts are used to associate absorbers with galaxies detected in the MUSE observations. }

\subsection{VLT/MUSE}
\label{sec:muse}

The \HI\ absorbers were targeted with MUSE in order to identify nearby galaxies and probe their ionised gas. MUSE is an integral-field spectrograph with a wavelength coverage of $\sim4800-9300~\AA$, providing spectroscopic redshifts {(see Sect. \ref{sec:lit_prop})}, strong emission lines and continuum images. {The spectral resolution is $R=1770$ at 4800 $\AA$ and  $ R= 3590$ at 9300 $\AA$, resampled to a spectral sampling of 1.25 $\AA$/pixel. } A total of 79 galaxies were identified within the MUSE field of view ($1'\times1'$, up to $\sim270$ kpc) and $\pm500$ km s$^{-1}$ of the absorbers, known as the primary MUSE sources. 

{Galaxies were identified through either their continuum emission or emission lines. ProFound \citep{Robotham:18} was used to identify continuum sources in the MUSE data. The varying exposure depths between fields result in different limiting magnitudes  \citep[$m_r=23.2-26.5$; see Table 1 of ][]{Weng:23a}, giving rise to non-uniform completeness for low-luminosity galaxies. Faint objects only detectable by their emission lines were systematically searched for without their continuum emission using the MUSE Line Emission Tracker (MUSELET) module of the MUSE Python Data Analysis Framework (MPDAF) package \citep{Bacon:16}. MPDAF was used to fit [\ion{O}{II}] $\lambda$$\lambda$3726, 3729, H$\delta$ $\lambda$4102, H$\gamma$ $\lambda$4340, H$\beta$ $\lambda$4861, [\ion{O}{III}]  $\lambda$$\lambda$4959, 5007, [\ion{N}{II}] $\lambda$$\lambda$6549, 6583, H$\alpha$ $\lambda$6563, and [\ion{S}{II}] $\lambda$$\lambda$6717, 6730 lines. The emission line flux limits range from $2.3\times10^{-18}$ ergs$^{-1}$ cm$^{-2}$ to $5.5\times10^{-17}$ ergs$^{-1}$ cm$^{-2}$ and can be found in Table 2 in \cite{Peroux:22}.  In addition, the deeper detection limits of the HST broadband imaging were used to identify source positions for spectral extraction from the MUSE data \citep[see Table 4 of][for the magnitude limits and corresponding central wavelengths]{Peroux:22}. This approach enabled the identification of ten additional [\ion{O}{II}] or Ly$\alpha$ emitters that were not detected by MUSELET. Finally, quasar PSF subtraction was applied to the MUSE observations, enabling the identification of galaxies at small projected separations from the quasar sightline.}

The $\pm500~\mathrm{km~s^{-1}}$ velocity criterion was adopted for consistency with the literature, and because recent studies have shown that {\ion{Mg}{ii}} absorbers, which trace dense neutral gas, are rarely associated with galaxies beyond $300~\mathrm{km~s^{-1}}$ \citep{Huang:21}.  {Twenty-seven of the 32 absorbers have galaxies identified at the absorber redshift within this velocity cut (found in 16 of the 19 MUSE fields), with 17 of these absorbers associated with multiple galaxies.}

\subsection{ALMA}

The galaxies were observed with ALMA to target CO emission lines, allowing us to characterise their molecular gas content. Thirty-nine of the galaxies were observed as part of the ALMA Large programme \citep[2023.1.00127.L, Cycle 10, PI: C. Péroux;][]{Bollo:26,Peroux:26}, with a further 21 covered by data previously published by the MUSE-ALMA Haloes team \citep{Klitsch:18,Kanekar:18,peroux:19,Szakacs:21}. We obtained 13 CO detections and 47 non-detections, while 19 sources were not observed. 

{The large programme targeted CO emission lines ( CO(2-1), CO(3-2), CO(4-3) ) using Band 4 (125 - 163 GHz) and Band 6 (211 - 275 GHz). To maximise the schedulability of the programme, the allowed range of spatial resolution was set to 0.57-1.02 arcsec, which resulted in observations being carried out in configurations C-2, C-3 or C-4. Data reduction was performed using the standard ALMA pipeline and CASA (version 6.5.6), reaching an rms sensitivity of 0.16 mJy beam$^{-1}$ over 50 km s$^{-1}$.  }

\subsection{{The primary sample}}

{The final primary sample consists of 79 galaxies within the MUSE field of view ($1'\times1'$, up to $\sim270$ kpc) and 500 km s$^{-1}$ of an \HI\ Lyman-$\alpha$ absorption feature in a background quasar spectrum. There are 19 background quasars sightlines, along which 32 absorbers were identified with HST, five of which had no nearby galaxies identified with MUSE. Seventeen absorbers have multiple associated galaxies, reflecting how the absorption does not necessarily probe the gas profile of a single galaxy, but can instead arise from the CGM of a gas-rich galaxy overdensity. The large number of absorbers with multiple associated galaxies suggests that the sample is intrinsically related to group environments and extended gaseous reservoirs, rather than isolated galaxies \citep{Weng:23a}. }

{The sample is primarily selected through \HI\ absorption, but several secondary selection effects should also be considered. The volume probed around each quasar sightline is limited by the MUSE field of view, with the corresponding physical area varying with redshift. The $\pm500$ km s$^{-1}$ cut may exclude additional members of the galaxy groups, an effect explored further in Sect. \ref{sec:groupmass} and Appendix \ref{appen:groups}. Furthermore, the identification of associated galaxies is limited by either emission-line flux or broad-band magnitude, with varying detection limits between the quasar fields. This biases the sample towards brighter and more actively star-forming galaxies. Finally, the \HI\ absorption may not be directly associated with every galaxy in the final sample. While these secondary selection effects impact the completeness of the sample, the analyses presented in this paper do not rely on statistics that require a complete sample, and therefore these effects do not affect our main conclusions. }

%%%%%%%%%%%%%%%%%%%%%%%%%%%%%%%%%%%%%%%%%%%%%%%%%%%%%%%

\section{Inferred quantities}
\label{sec:quantities}

In this study we build on on the work done by the MUSE-ALMA Haloes team to characterise the baryons in the galaxy sample. Here we summarise the key galaxy properties derived in other studies (stellar mass, star formation rate, metalicity, molecular gas mass, and kinematic and morphological models) and outline additional quantities estimated for this analysis (expected \HI, dynamical and group masses; these are available in Appendices \ref{appen:mega_plot} and \ref{appen:groups}). Due to the evolving nature of the MUSE-ALMA Haloes {programme}, as well as varying sensitivities and resolutions, all properties are available for a fraction of the 79 galaxies. {A complete set of stellar mass, \HI\ mass, H$_2$ mass (or an upper limit), and dynamical mass measurements is available for 27 galaxies.} Figure \ref{fig:mega_plot} in Appendix \ref{appen:mega_plot} summarises the properties available for each galaxy.

\subsection{Properties derived in previous MUSE-ALMA Haloes studies}
\label{sec:lit_prop}

{Galaxy redshifts were measured from the MUSE spectra following the procedure described in \citet{Peroux:22}. {Redshifts were determined using the MARZ software \citep{Hinton:16} for galaxies identified using the methods outlined in Sect.~\ref{sec:muse}.  The MARZ software} combines a visualisation tool and a template cross-correlation tool for each source with quasar, galaxy and stellar template spectra.  Of the 79 galaxies, 73 have robust redshifts determined from high-S/N emission and absorption lines, while the remaining six have redshift measurements based on low-S/N emission or absorption features.  }

\cite{Weng:23a} calculate the star formation rate (SFR) following the \cite{Kennicutt:98} prescription, using H$\alpha$ emission line where available, or the [\ion{O}{II}] luminosity.  Additionally, \cite{Weng:23a} calculate the galaxy metallicities {for emission line detected galaxies} using the $R_3 = \log(F([$O {\sc iii}$]\lambda5007)/F($H$\beta))$ auroral strong-line calibration \citep{Curti:17}. For sources {with} $z<0.4$, O$_3$N$_2$ \citep{Pettini:04} is measured to remove the degeneracy of the double branched $R_3$ indicator. {Halley et al. (in prep) provide metallicities of the cool gas probed in absorption, corrected for ionisation and dust depletion effects, along with the line-of-sight depletion factors.}

\cite{Augustin:24} derive stellar masses from SED fitting with {\tt LE PHARE} \citep{Arnouts:99,Ilbert:06} code.  {For the fitting, they primarily use HST magnitudes, adopting photometry from DECaLs when HST magnitudes are unavailable.} {The available photometric bands cover the spectral region around the 4000 \AA\ break, enabling constraints on stellar masses from SED fitting.} Robust stellar mass estimates based on {detections in} $2-3$ bands are obtained for 34 galaxies, while a further 19 galaxies have `probable' estimates derived from a {detection in a single band, and upper limits in the other bands}. {While most fields are covered by three HST bands, galaxies are not always detected in all of them. Non-detections are most common in the bluest band, particularly for faint galaxies that were identified in MUSE solely through their emission-line signatures and exhibit little or no detectable continuum emission.} {Due to contamination from a nearby companion,} Q0454-220$\_$4 has been reanalysed and its stellar mass was revised from $\log(M_*)=12.4\pm0.1$ to $\log(M_*)=10.7\pm0.1$. Throughout the figures of this paper, MUSE-ALMA Haloes sources with robust stellar masses are shown by markers with black outlines, while those with `probable' stellar masses have no outline.

\cite{Bollo:26} calculate the molecular hydrogen gas mass ($M_{H_2}$) from the CO luminosity using a metallicity-dependent conversion factor ($\alpha_{CO}$;  the geometric mean of the prescriptions proposed by \citeauthor{Bolatto:13}~\citeyear{Bolatto:13} and \citeauthor{Genzel:12}~\citeyear{Genzel:12}). For sources without a direct metallicity measured from the MUSE data, the stellar mass-metallicity relation is assumed \citep{Maiolino:08,Genzel:15}. For the {eight} galaxies {for which neither} a direct metallicity measurement {nor} a reliable stellar mass {is available}, a value of $\alpha_{CO}\sim15$ is adopted  {based on the low metallicity of the gas in absorption}. 

\cite{Weng:23b} create kinematic models from the MUSE data using {\tt GALPAK$^{\rm 3D}$} \citep{Bouche:15}.   This provides estimates of galaxy parameters including inclination, kinematic position angle and maximum rotation velocity.  {\cite{Barfety:26} examine the cold molecular gas distribution and kinematics of a subset with sufficiently high resolution and signal-to-noise ratio.} \cite{Karki:23} derive morphological models by fitting Sérsic profiles to the reddest HST images using {\tt GALAPAGOS} \citep{Haussler:11}. This provides morphological parameters such as effective radius, morphological position angle, axis ratio, and Sérsic index.

\subsection{Estimating \HI\ masses}
\label{sec:MHI}

To study the distribution of baryons in the MUSE–ALMA Haloes galaxies, it is important to investigate their \HI\ content. The \HI\ gas acts as reservoir in the baryon cycle, connecting the accretion of gas from IGM and CGM to the formation of dense, star-forming, molecular gas. Since the MUSE-ALMA Haloes galaxies were selected by their proximity to \HI\ absorbers, they are likely to host significant \HI\ reservoirs. The available multiwavelength observations of the galaxies, along with the known absorber column densities and impact parameters, offer a valuable opportunity to explore the \HI\ content of the galaxies and connect absorption properties to galaxy-scale gas reservoirs. In this section, we calculate the \textit{estimated} \HI\ masses of the galaxies based on their multiwavelength properties. Although caution is required when interpreting results for individual galaxies, the estimated \HI\ masses are useful for characterising the MUSE–ALMA Haloes sample as a whole.

 {First we perform a number of photometric corrections to the optical magnitudes measured with MUSE.} We account for the galactic extinction of the Milky Way, which reddens the light incoming from galaxies. We apply the following correction: $ m_{x,GEC} = m_x - R_x \times E(B-V)$. We use the $R_x$ values provided in \cite{Schlegel:98} and the $E(B-V)$ values from \cite{Schlafly:11}. Next, we apply a k-correction to convert the magnitudes measured in the observed filter to corresponding rest-frame band. For the MUSE data, we have observed magnitudes in the $r$, $i$, $V$ and $R$-bands. These magnitudes, along with the galaxy redshift, are input into the {\tt kcorrect} code \citep{Blanton:07}  which maps these observed-band magnitudes to the rest-frame $g$ and $r$-bands.

 {To estimate the \HI\ masses of the MUSE-ALMA Haloes galaxies, we adopt the method from \cite{Zu:20}, which uses the stellar mass and $g-r$ colour.} Using ALFALFA \HI\ data and SDSS optical data, their statistical method accounts for Malmquist bias by incorporating the ALFALFA detection probability for each SDSS galaxy. This approach avoids overestimating the \HI\ to stellar mass ratio for galaxies not detected by ALFALFA, without requiring a trade-off between survey volume and depth. Their best fitting relation is given by equation \ref{eq:MHI}:
\begin{equation}
    \frac{M_{\rm HI}}{M_*} = -0.328 \log(\frac{M_*}{h^{-2}\rm M_{\odot}}) - 1.492(g-r) + 6.662,
    \label{eq:MHI}
\end{equation}
where $M_{\rm HI}$ is the \HI\ mass, $M_*$ is the stellar mass, and $(g-r)$ is the galaxy colour. This relation has a scatter of $\sigma = 0.272$~dex, comparable to that of other estimators discussed in Appendix~\ref{appen:HI}. The scatter is dominated by the intrinsic scatter arising from the numerous physical processes that build and deplete the gas reservoir, including the galaxy chemical evolution and environmental effects. To confirm the accuracy of the method, \cite{Zu:20} compare their \HI\ to stellar mass ratio estimations with measured values from the GALEX Arecibo SDSS Survey \citep[GASS;][]{Catinella:10}. They use a stellar mass dependent colour cut to compare blue and red galaxies \citep{Zu:16}: $(g-r)_{\rm cut} = 0.8(\frac{\log(M_{*}/h^{-2}\rm M_{\odot})}{10.5})^{0.6}$. \cite{Zu:20} find that their relation is able to recover the measured \HI\ to stellar mass ratio well for both blue and red galaxies, although the scatter is larger for red galaxies ($\sigma_{\rm blue}=0.300$, $\sigma_{\rm red}=0.375$). Applying this same blue/red cut, $72\%$ of galaxies in the MUSE–ALMA Haloes sample are classified as blue,  {where the relation is more robust}. 

 {This \HI\ mass estimator is calibrated at $z\sim0$ and applied here at $z\sim0.5$. This is justified by the expectation that the cosmic \HI\ density evolves only weakly over this redshift range \citep{Peroux:20}. Consistent with this, \HI\ stacking studies indicate that correlations between optical galaxy properties and \HI\ mass at $z\sim1$ are broadly consistent with those observed locally \citep{Chowdhury:24}. Theoretical models also support this picture: semi-analytic simulations such as SHARK \citep{Lagos:18} predict a relatively flat evolution in the global \HI\ density out to $z\sim1.5$. At higher redshifts, these models begin to diverge from observations, likely reflecting the increasing contribution of the circumgalactic medium, where a larger fraction of the neutral gas reservoir is expected to reside.
} 

Using this method, we are able to estimate the \HI\ to stellar mass ratio, and subsequently the \HI\ mass, for 50 out of the 79 MUSE-ALMA Haloes galaxies. {Faint galaxies, particularly those identified only via emission-line detections, lack sufficient continuum emission for $g-r$ colour measurements and SED-based stellar mass estimates, and consequently do not have \HI\ mass estimates. } Figure \ref{fig:MHI_Mstar} shows the estimated \HI\ mass as a function of stellar mass for the MUSE-ALMA Haloes galaxies along with measurements from the $z\sim0$ xGASS sample \citep{Catinella:18}, the absorber-selected $z<0.14$ sample from \citet{Kulkarni:22}, and the \HI\ stacking analyses from MIGHTEE-\HI\ \citep{Sinigaglia:24} at $z\sim0.4$ and CATz1 \citep{Chowdhury:22} at $z\sim1$.   Within the overlapping stellar mass range, the MUSE-ALMA Haloes galaxies  {have estimated \HI\ masses that overlap with, and bridge, the $z\sim0$ and $z\sim1$ samples.}

\begin{figure}[h!]
    \centering
    \includegraphics[width=\linewidth]{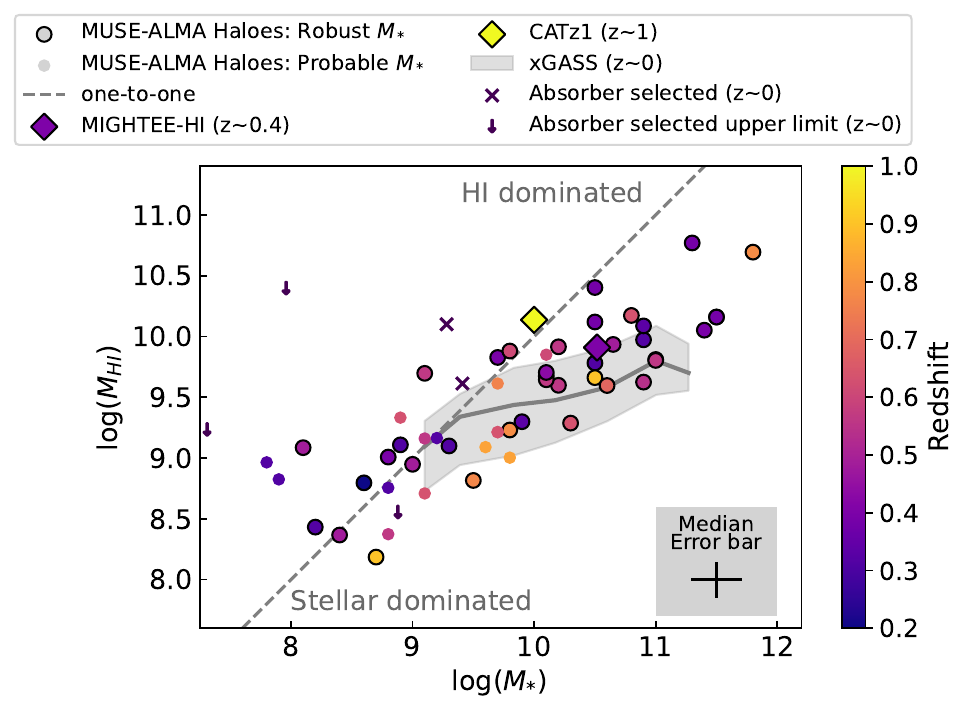}
    \caption{Estimated \HI\ mass as a function of stellar mass. The MUSE-ALMA Haloes galaxies are shown as circles, coloured from yellow to purple by their redshift.  {The diamond markers show the \HI\ stacking results from \cite{Sinigaglia:24} at $z\sim0.4$ and \cite{Chowdhury:22} at $z\sim1$. The grey shaded region and solid line shows the interquartile range and rolling median of the ($z\sim0$) xGASS sample for comparison. The `$\times$' markers and arrows show the absorber-selected $z<0.14$ sample from \cite{Kulkarni:22} with 21cm \HI\ measurements.}  }
    \vspace{-5pt}
    \label{fig:MHI_Mstar}
\end{figure}

\subsection{Dynamical masses}
\label{sec:MD}

We combine the results from the kinematic and Sérsic modelling presented in \cite{Weng:23b} and \cite{Karki:23} to estimate the total  {dynamical} masses of the MUSE–ALMA Haloes galaxies. The total dynamical mass ($M_{Dyn}$) enclosed within radius $R$ is given by $M_{Dyn}(<R) = Rv_{\rm circ}^2 /G,$ where $v_{\rm circ}$ is the circular velocity and $G$ is the gravitational constant.  {The circular velocity can be derived from the rotation velocity with an asymmetric drift correction to account for pressure support \citep{Burkert:10}. For an exponential disk, $v_{\rm circ}^2 = v_{\rm rot}^2 + 2(R/R_s)\sigma^2$, assuming $R_s = R_e/1.678$, {where $R_e$ is the effective radius from Sérsic fits to HST imaging by \cite{Karki:23}}.  We adopt the maximum rotational velocity and velocity dispersion derived from the kinematic models of the ionised gas emission lines from MUSE \citep{Weng:23b} as the rotation velocity ($v_{\rm rot}$) and velocity dispersion ($\sigma$) respectively.} {For reliable kinematic modelling, \cite{Weng:23b} adopted a signal-to-noise ratio cut of {S/N} $>3$ and required galaxies to satisfy  $R_e/R_{\rm PSF}>1.5$.} As the ionised gas emission traces the inner regions (in contrast to the extended \HI\ discs in most nearby star-forming galaxies), we compute the dynamical mass within a comparable extent, adopting $R = 2R_e$.

To estimate the uncertainty in the dynamical mass, we assume an uncertainty of 20 km s$^{-1}$ in the circular velocity and 0.25 kpc in the effective radius, and propagate this to the dynamical mass. Although the kinematic and Sérsic models provide formal uncertainties on the rotation velocity and effective radius, we adopt these larger values to more realistically capture the overall uncertainty. In addition to these uncertainties, we emphasise that these total mass estimates are valid only for rotation-supported, self-gravitating systems.  Complementary constraints from molecular gas kinematics provide an independent check on these dynamical masses.  Models for four CO detections  {using} high resolution ALMA data have been constructed by \cite{Barfety:26}. As our analysis focuses on the global sample rather than detailed modelling of individual systems, some discrepancies are expected due to the different methods used.  {The dynamical masses derived from the ALMA data are in broad agreement with those from the MUSE data: all four sources agree within 0.4 dex, with two consistent to within 0.1 dex. For consistency across the sample, we adopt the MUSE results for these four sources.}

\subsection{ {Overdensity} masses}
\label{sec:groupmass}

In addition to estimating the dynamical mass within $2R_{e}$ for each galaxy, we also derive masses for overdensities in which three or more galaxies reside. As these absorber systems are not robustly defined groups, we refer to them as galaxy overdensities. We follow the method adopted in the previous MUSE-ALMA Haloes study by \cite{Weng:23a}, estimating the group masses using the virial theorem: $M_{\rm group} \sim 3 R \sigma_{\rm group}^2 / G$, where R is defined as half the maximum separation between galaxies in the overdensity, $\sigma_{\rm group}$ is the velocity dispersion, and $G$ is the gravitational constant. The velocity dispersion is calculated using the \textsc{gapper} function \citep{Beers:90}, which is well suited to systems with a small number of members. We then apply a correction to account for the presence of a central galaxy in each group, which is assumed to be at rest with the underlying dark matter halo, following \cite{Yang:07}. The \textsc{gapper} estimator and the corrected velocity dispersion are given by Equation \ref{eq:gapper}: 
\begin{equation}
    \sigma_{\rm group} = \frac{\sqrt{\pi}}{N(N-1)}  \sum^{N-1}_{i=1} w_ig_i, ~~~~~ \sigma_{\rm group, corr} = \sqrt{\frac{N}{N-1}} \sigma_{\rm group} ,
    \label{eq:gapper}
\end{equation}
where $N$ is the number of galaxies in the group, $w_i$ is the weight defined as $w_i = i(N-i)$, and $g_i$ is the gap in ordered velocities of the galaxies ($g_i = v_{i+1}-v_i$). We approximate the uncertainty in the velocity dispersion as the MUSE velocity resolution at 7000~\AA\ (54 km s$^{-1}$) and propagate it to estimate the uncertainty in the group mass.

We revise the masses calculated by \cite{Weng:23a} using the 79 primary galaxies in the MUSE-ALMA Haloes {programme}. As noted in their work, these group masses are likely lower limits because many systems may contain objects outside of the MUSE field of view {and the 500 km s$^{-1}$ velocity requirement}. Consequently, we also estimate upper limits to the group masses by including galaxies {with spectroscopic redshifts from MUSE up to 1000 km s$^{-1}$,} and within the HST field of view (up to 546 kpc) that have photometric redshifts \citep{Yang:23} and lie within 1000 km s$^{-1}$ of the absorber. {Appendix \ref{appen:groups} provides a table listing the lower and upper limits on the group masses (Table \ref{tab:groups}). It also discusses the distribution of velocity separations between associated galaxies and absorbers, as well as the impact of the maximum velocity cut.}  {Group masses are calculated for 11 absorber systems containing three or more galaxies from the primary MUSE sample (includes 56/79 galaxies) and a total of 16 absorber systems containing three or more galaxies after including the supplementary MUSE and HST targets. }

%%%%%%%%%%%%%%%%%%%%%%%%%%%%%%%%%%%%%%%%%%%%%%%%%%

\section{Baryon Reservoirs}
\label{sec:baryon_res}

Characterising the amount and distribution of the different gas phases within the ISM and CGM provides crucial insight into the evolution of the galactic baryon cycle. In this section, we {establish a census of the condensed baryons in the MUSE-ALMA Haloes galaxies {(the 79 galaxies in the primary MUSE sample)}. As an \HI\ absorber-selected sample, they span a remarkably wide stellar mass range of $\sim4$ orders of magnitude, $7.8<\log(M_*/\rm M_{\odot})<11.8$. We compare their properties with other samples spanning a range of redshifts and selection criteria, in order to build a more complete picture. 

Galaxy samples with direct measurements of both CO and \HI\ are valuable for providing consistent comparison datasets. The local Universe serves as an ideal benchmark due to the sensitivity and completeness of available data. The xCOLD GASS and xGASS surveys \citep[][]{Saintonge:17,Catinella:18} and ALLSMOG \citep{Cicone:17} provide complementary CO and \HI\ measurements for galaxies in the local Universe. They show that gas fractions increase with decreasing stellar mass, and that \HI\ dominates over molecular gas irrespective of stellar mass. The CHILES survey \citep[][]{Fernandez:16,Hess:25,Bluebird:26} and HIGHz survey \citep[][]{Cortese:17}  provide CO and \HI\ observations at intermediate redshift $z\sim0.1-0.5$. They find evidence to suggest that the ratio of molecular to atomic gas increases with increasing redshift. Additionally, \cite{Jarvis:25} report 11 \HI\ detections at $z\sim0.25$–0.35 from the MIGHTEE-\HI\ survey. Using these data, they provide the first extension of the baryonic Tully–Fisher relation (baryonic mass as a function of emission line width) beyond $z=0.25$. They find broad consistency with local measurements, but tentative evidence for flattening in the high mass regime, which may indicate evolution in the \HI\ content with redshift or an increasing contribution from molecular gas.

When considering only CO, it is possible to extend studies to higher redshifts. The PHIBSS surveys \citep{Tacconi:13} target the molecular gas content of star-forming galaxies on and around the star-formation main sequence around cosmic noon.  In particular, PHIBSS-2 \citep{Freundlich:19} includes 60 CO detections at $0.5 < z < 0.8$, providing significant overlap with the MUSE-ALMA Haloes sample. The PHIBBS surveys demonstrate that the molecular gas fraction increases between the local Universe and cosmic noon.

{For the MUSE-ALMA Haloes sample, we adopt a \cite{Chabrier:03} initial mass function (IMF), while the literature comparison samples adopt either \cite{Chabrier:03} or \cite{Kroupa:01} IMFs. The resulting differences in stellar mass estimates are small \citep[$\sim0.03$ dex;][]{Madau:14} and are not expected to significantly affect the comparison. MUSE-ALMA Haloes, xGASS, and PHIBSS all use metallicity-dependent $\alpha_{\rm CO}$ prescriptions to estimate $M_{\rm H_2}$, whereas the remaining comparison samples assume a fixed Galactic $\alpha_{\rm CO}$. The choice of $\alpha_{\rm CO}$ can therefore introduce systematic offsets in the derived $M_{\rm H_2}$ values and molecular gas fractions.}

\subsection{The ISM}

\begin{figure}[h!]
    \centering
    \includegraphics[width=0.88\linewidth]{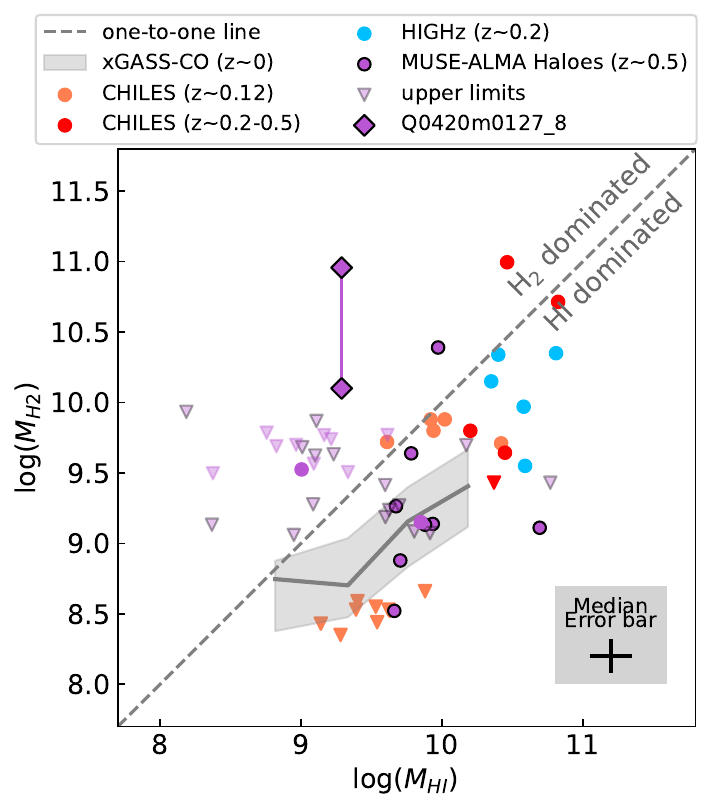}
    \caption{ H$_2$  mass against estimated \HI\  mass for the MUSE-ALMA Haloes galaxies (purple). {Black outlines indicate the MUSE-ALMA Haloes galaxies with the most robust stellar mass estimates. The two diamond markers represent Q0420m0127$\_$8, assuming metallicity-dependent and LIRG values of $\alpha_{\rm CO}$.}  The grey shaded region and solid line show the interquartile range and rolling median of the ($z\sim0$) xGASS-CO sample \citep{Saintonge:17,Catinella:18}, the blue points show the HIGHz ($z\sim0.2$) galaxies \citep{Cortese:17}, and the orange and red points {show} the CHILES galaxies,  \citep{Fernandez:16,Hess:25,Bluebird:26}.  } 
    \vspace{-5pt}
    \label{fig:MHIMH2}
\end{figure}

Figure \ref{fig:MHIMH2} presents the estimated \HI\ masses and the measured H$_2$ masses (or upper limits) inferred from the CO observations for the MUSE-ALMA Haloes galaxies, together with the comparison samples.  Across these different samples, most galaxies are found to host either more \HI\ than H$_2$, or comparable amounts of each, as expected.  {\HI\ serves as the reservoir from which molecular gas forms, which must be shielded from radiation to remain cold and dense. The molecular gas is continually consumed by star formation and can be dissociated back into \HI\ by stellar feedback \citep{Krumholz:09,Saintonge:22}. Due to observational limitations, many studies approximate the total cold gas mass as $M_{\rm gas} \simeq M_{\rm mol}$ \citep[e.g.][]{Carilli:13,Wuyts:16,Suzuki:21}. We emphasise here that \HI\ does constitute a significant baryonic component  {of the ISM when considering scales beyond the inner stellar disc,} and is expected to evolve only gradually with redshift \citep{Peroux:20,Walter:20}. Its contribution should not be overlooked at intermediate and high redshifts, motivating our adoption of an indirect \HI\ estimation in our analysis. Using this approach, } the MUSE-ALMA Haloes galaxies broadly follow this trend, hosting either more \HI\ than H$_2$ or comparable amounts of each, with one notable exception: Q0420m0127$\_$8, whose inferred H$_2$ mass lies 1.7~dex above its estimated \HI\ mass and 0.7~dex above its stellar mass. 

\begin{figure*}[h!]
    \centering
    \includegraphics[width=0.8\linewidth]{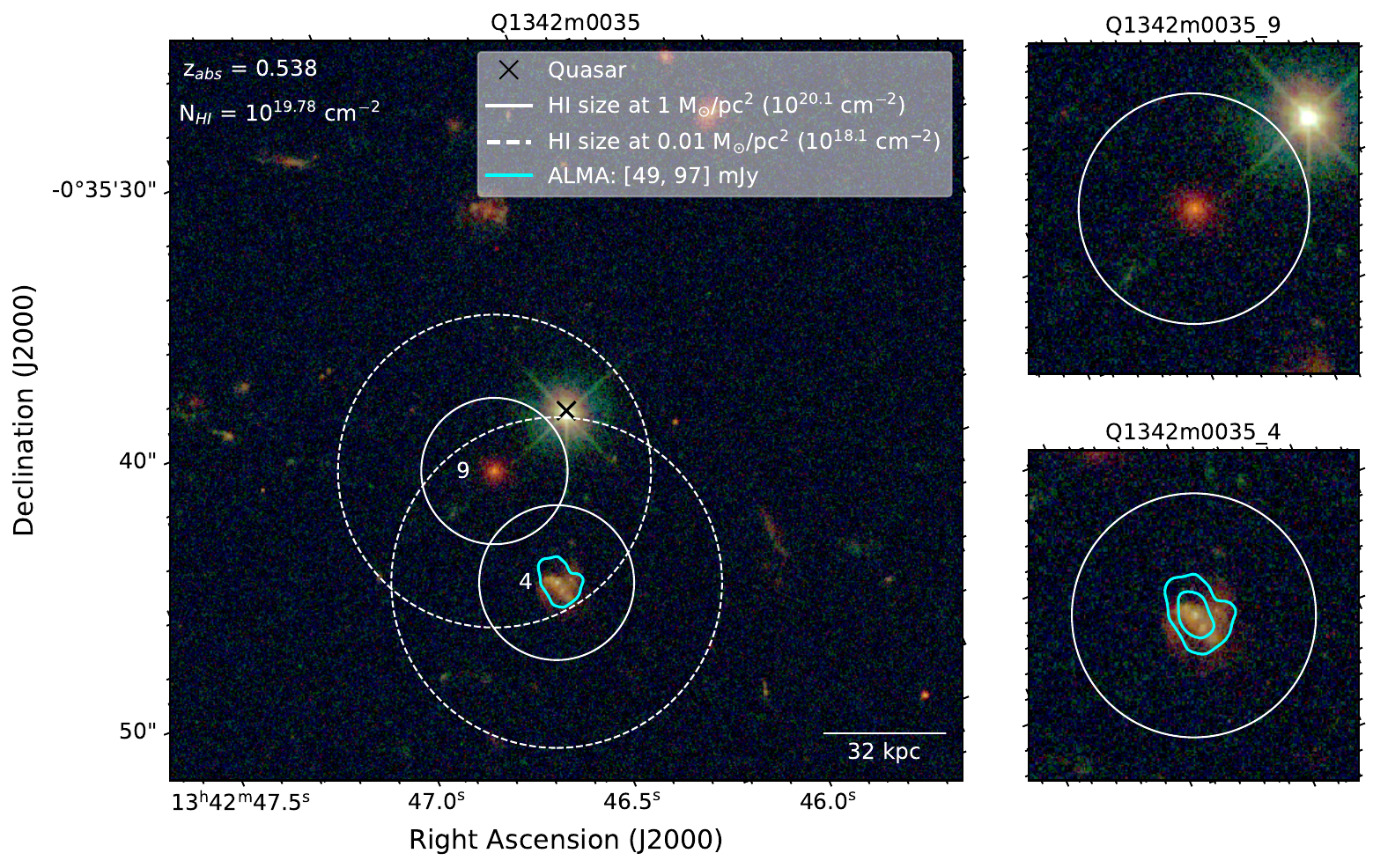}
    \caption{The quasar field Q1342–0035.  The background is a three-colour image constructed from the HST F336W, F475W, and F814W bands. This quasar is located at the centre of the image and has one absorber and two nearby galaxies at $z\sim0.54$. {The \HI\ sizes are inferred from the \HI\ size-mass relation, using the optically derived \HI\ mass.}   }
    \vspace{-5pt}
    \label{fig:overlay1342}
\end{figure*}

Q0420m0127$\_$8 was studied by \cite{Klitsch:18}. Adopting the metallicity-dependent $\alpha_{CO}=4.5$ from \cite{Bollo:26} yields a high molecular gas mass of $M_{\rm H2}=10^{10.95}$ M$_{\odot}$. On the other hand, \cite{Klitsch:18} showed that the ratio of $L'_{CO(3-2)}$ and $L'_{CO(2-1)}$ provides evidence that this source is a luminous infrared galaxy (LIRG), a very actively star forming and dusty galaxy. They therefore adopt a lower conversion factor, $\alpha_{CO}=0.6$, which is appropriate for LIRGs \citep{Papadopoulos:12b,Papadopoulos:12a}, resulting in a lower, more typical molecular gas mass of $M_{\rm H2}=10^{10.1}$ M$_{\odot}$. This galaxy, together with both of its inferred H$_2$ masses, is indicated by diamond symbols in Fig.~\ref{fig:MHIMH2} and in subsequent figures.

\subsection{The CGM and local environment}

We now turn to the gas in the surroundings of the galaxies, that is, the CGM.  {Over $50\%$ of the \HI\ absorbers have two or more nearby galaxies within  {0.5} arcmin (up to $\sim270$ kpc) and 500 km s$^{-1}$ of the absorber. Figure~\ref{fig:overlay1342} shows one of the quasar fields (Q1342–0035) as an example. This quasar is located at the centre of the image and has one absorber and two nearby galaxies at $z\sim0.54$. The \HI\ radii at 1 M$_{\odot}$ pc$^{-2}$ ($10^{20.1}$ cm$^{-2}$) and 0.01 M$_{\odot}$ pc$^{-2}$ ($10^{18.1}$ cm$^{-2}$), inferred from the \HI\ size-mass relationships \citep{Wang:16,Wang:25}, are shown by the solid and dashed circles respectively. This demonstrates the overlapping neutral gas reservoirs of the two galaxies and the quasar sightline. One galaxy was detected in CO by ALMA, with  {$3\sigma$ and $6\sigma$} contours overlaid in cyan, highlighting how molecular gas is typically more centrally concentrated than atomic gas. The remaining absorber-galaxy systems are presented in Appendix \ref{appen:groups}. } 

As described in Sect. \ref{sec:groupmass}, we estimate group masses for overdensities with at least three members in either the primary MUSE sources or the extended field (within the MUSE and HST fields of view and 1000 km s$^{-1}$), i.e., {16} absorber systems. {Ten} of these have five or fewer members. As such, our overdensities span a narrow dynamic range of $10^{12}$-$10^{13.8}$ M$_{\odot}$, with the exception of one low mass system (Q1211p1030, $z_{\rm absorber}$ =0.3929, { $M_{\rm group}=1\times10^{10}$ M$_{\odot}$).} A table summarising the absorber systems can be found in Appendix \ref{appen:groups}.

We investigate the effect of overdensity size on the properties of individual galaxies.  {Environmental processes are known to affect both the gas content and star-forming properties of galaxies. More massive haloes typically host a higher fraction of passive, quenched, early-type systems as a result of external processes including tidal interactions, galaxy harassment and ram pressure stripping \citep[e.g.][]{VanKempen:24}. } However, we find no correlation between the group masses derived in Sect. \ref{sec:groupmass} and either the absorber column density or galaxy properties, including stellar mass, gas mass, SFR, metallicity  {and baryon fraction}, {regardless of whether the upper or lower group mass estimates are used}. We note that the overdensities are not robustly defined groups, and most have few members. Given the limited dynamic range in group masses, together with the heterogeneous nature of our sample and large uncertainties, the absence of clear trends is perhaps  {expected}. {More complete spectroscopic redshift coverage outside the MUSE field of view and beyond $\pm500$ km s$^{-1}$, together with improved coverage of passive galaxies lacking high-S/N emission or absorption features, would improve group membership assignments, potentially increasing the dynamic range of group masses, and help constrain any underlying galaxy-group trends.} {A discussion of the maximum velocity cut is presented in Appendix \ref{appen:groups}.}

Additionally, we investigate the connection between the \HI\ absorber properties and the galaxy properties.  {\citet{Klimenko:23} examine \HI\ column density ($N_{\rm HI}$) as a function of impact parameter ($b$; the projected distance between the absorber and the galaxy) for absorber-selected galaxies across a range of redshifts, including the MUSE-ALMA Haloes sample, and compare with simulations. They find a general decline of $N_{\mathrm{HI}}$ with increasing $b$, good agreement between observations and simulations, and a small number of outliers at $z>0.7$ in the MUSE-ALMA Haloes sample with elevated $N_{\mathrm{HI}}$.}

 {We compare the \HI\ masses derived from the galaxy optical properties to \HI\ masses inferred from the CGM.} { \cite{Wang:26} combine deep \HI\ observations from FEASTS, a single-dish survey conducted with FAST \citep{Jiang:19}, with high resolution interferometric data from a range of surveys and archival data, enabling a detailed study of \HI\ in and around local gas-rich galaxies. Using {these} data, they investigate the extent of \HI\ around associated galaxies as a function of column density limit. They empirically derive cumulative probability distributions of the impact parameter relative to the galaxy centre for \HI\ at different column densities.  They find a correlation between the ratio of $b$ and $r_{001}$ (the radius at which the \HI\ has a surface density equal to 0.01 M$_{\odot}$ pc$^{-2}$), with $N_{\rm HI}$, for column densities $\log(N_{\rm HI}/ \rm cm^{-2})<20.5$. This relationship is given by Equation \ref{eq:r001_NHI}:
\begin{equation}
    b/r_{001} =\log(N_{\rm HI}/ \rm cm^{-2})\times(-0.19\pm0.01)+(4.26\pm0.25).
    \label{eq:r001_NHI}
\end{equation}
Using our measured impact parameters and column densities, we estimate the corresponding $r_{001}$ values and then apply the \HI\ size–mass relation from \cite{Wang:25} to infer the \HI\ mass.  {The $1\sigma$ scatter of this estimate, considering both the scatter of $b/r_{001}$ at a given $N_{\rm HI}$ and the size mass relation, is 0.12 dex \citep{Wang:26}.} A comparison between the \HI\ masses derived using this method and those estimated from optical properties (Sect.~\ref{sec:MHI}) is shown in Fig.~\ref{fig:MHI_MHI}. 

Although both approaches provide indirect estimates of the \HI\ mass, we find some agreement for $b\lesssim 100$~kpc, approximately the maximum impact parameter probed in the \cite{Wang:26} study. At larger impact parameters this relation is likely  {less robust}. {Twenty-one ({eight}) sources have \HI\ masses (upper limits) estimated from the absorber properties for absorbers with $b < 100$~kpc and $\log(N_{\mathrm{HI}}/\mathrm{cm}^{-2}) < 20.5$.} {Additionally, the group environment plays a role, as massive galaxies closer to the absorber can influence the other galaxies. For example, in the system shown in Fig. \ref{fig:overlay1342}, the source closest to the absorber, Q1342m0035$\_9$, exhibits a small offset in \HI\ masses of only 0.2 dex. In contrast, Q1342m0035$\_4$, located further from the quasar sightline and whose \HI\ profile overlaps with the nearer galaxy, has a larger offset of 0.7 dex. } Sources whose \HI\ profiles overlap with the quasar sightline, and do not overlap with that of a closer galaxy, are highlighted with a green outline in Fig. \ref{fig:MHI_MHI}. We refer to these as the `dominant' galaxies, and they lie closest to the one-to-one relation. {For non-dominant galaxies with $b \lesssim 100$ kpc, the absorber-based \HI\ masses tend to lie above the one-to-one relation, suggesting that the method overestimates their \HI\ masses because of the enhanced $N_{HI}$ from the dominant galaxy. }

While the $1\sigma$ scatter of the $N_{\rm HI}$-based method is smaller than the optically-based method, we note that the irregular \HI\ morphology and environmental effects are likely to significantly influence the estimated \HI\ mass. Thus we adopt the optically-derived \HI\ masses for the remainder of the results. Nevertheless, the $N_{\rm HI}$-based method provides a valuable, independent means of estimating \HI\ masses without relying on optical properties, particularly for high-redshift galaxies where \HI\ emission cannot yet be directly observed.

\begin{figure}[h!]
    \centering
    \includegraphics[width=0.9\linewidth]{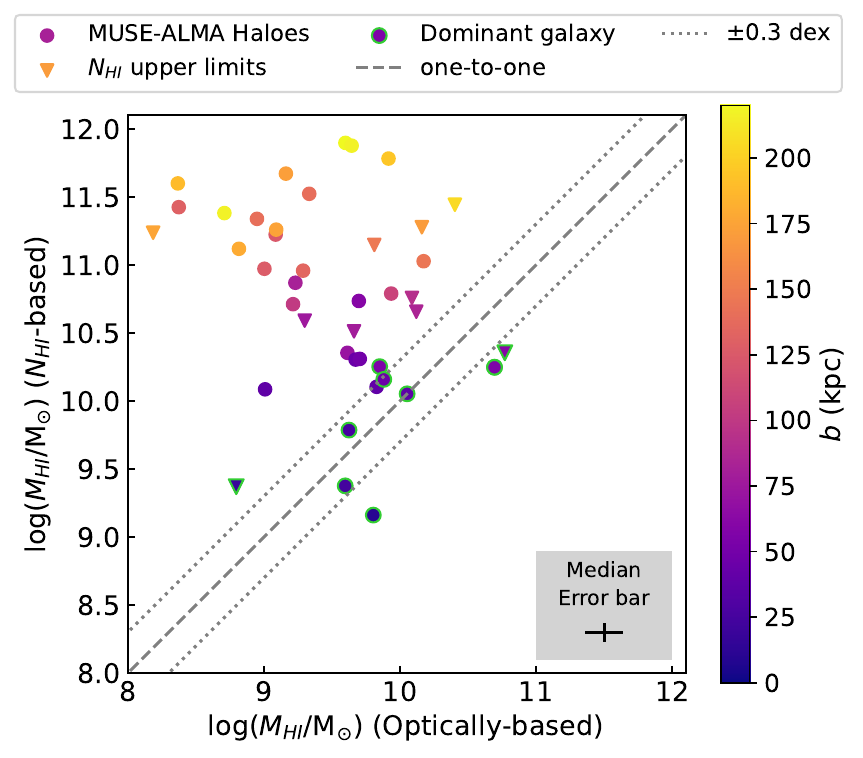}
    \caption{ {Comparison of \HI\ masses estimated using scaling relations of local resolved galaxies. \HI\ masses are estimated using optical properties following the method of \cite{Zu:20}, which uses $g-r$ and $M_*$, and using \HI\ absorber properties following the method of \cite{Wang:26}, which uses $N_{\rm HI}$ and $b$.} }
    \vspace{-10pt}
    \label{fig:MHI_MHI}
\end{figure}

\section{Baryon content and fractions}
\label{sec:fractions}

\subsection{Cold gas fractions}
\label{sec:gas_frac}
Examining the gas fractions is an effective way to characterise how the different components of galaxies contribute to their overall baryon budget. Figure \ref{fig:fgas} shows the \HI\ to stellar mass ratios, and H$_2$ to stellar mass ratios as a function of stellar mass in the upper and lower panels respectively.  We find that low-mass galaxies are  {the bluest and most} gas-rich.  The largest \HI-to-stellar mass ratio {of a MUSE-ALMA Haloes galaxy with a robust stellar mass estimate is $M_{\rm HI}/M_{*}=10\pm4$, and the largest H$_2$-to-stellar mass ratio is $M_{\rm H2}/M_{*}=0.6\pm0.4$ }(source Q0420m0127$\_$8, assuming the LIRG $\alpha_{CO}$). As shown in Fig. \ref{fig:MHIMH2}, if the metallicity-dependent $\alpha_{CO}$ is adopted instead, Q0420m0127$\_$8 becomes very molecular gas-rich relative to its \HI\ mass and similarly is also an outlier with respect to stellar mass.  {We also note that adopting the alternative $N_{\mathrm{HI}}$-based method for estimating \HI\ mass does not alter our conclusion regarding \HI\ dominance in low-mass galaxies, as this approach generally predicts even higher \HI\ masses.} In contrast, massive galaxies are less gas-rich. This suggests that stellar mass builds up more rapidly than the cold gas reservoir. The lowest \HI-to-stellar mass ratio  {of a MUSE-ALMA Haloes galaxy with a robust stellar mass estimate is  $M_{\rm HI}/M_{*}=0.04\pm0.03$ and the lowest H$_2$-to-stellar mass ratio of $M_{\rm H2}/M_{*}=0.002\pm0.7$. }

We fit linear relations in log–log space to both the atomic and molecular gas fraction versus stellar mass plots. For the molecular gas fraction fit, we apply weighting based on upper limits using PyMC \citep{pymc2023}. The resulting fits are provided in Equations \ref{eq:fHI} and \ref{eq:fH2} respectively.} 
\begin{equation}
    \log \left( \frac{M_{\rm HI}}{M_*} \right) = (-0.50\pm0.05)\log \left(  \frac{M_*}{\rm M_{\odot}} \right) + (4.52\pm0.46).
    \label{eq:fHI}
\end{equation}
\begin{equation}
    \log \left( \frac{M_{\rm H2}}{M_*} \right) = (-0.88\pm0.08)\log \left(  \frac{M_*}{\rm M_{\odot}} \right) + (7.65\pm0.80).
    \label{eq:fH2}
\end{equation}
 {We note that the atomic gas fraction is intrinsically correlated with stellar mass (Eq.~1), with colour providing an additional important dependence. However, only seven dominant galaxies have both \HI\ masses estimated using the absorber-based method (with $b<100$ kpc and $\log(N_{\rm HI}/\rm cm^{-2})<20.5$) and measured stellar masses, providing too small a sample to robustly constrain the slope of this relation using an alternative method.} The steeper slope for molecular gas may suggest that the massive galaxies in the MUSE-ALMA Haloes sample consume their molecular gas more efficiently,  {while the more extended \HI\ reservoir is less affected, or that there is an inefficient conversion of \HI\ to H$_2$}. However, the small number of detections and the relatively ill-constrained upper limits for low-mass galaxies make this trend tentative.

\begin{figure}[h!]
    \centering
    \includegraphics[width=0.8\linewidth]{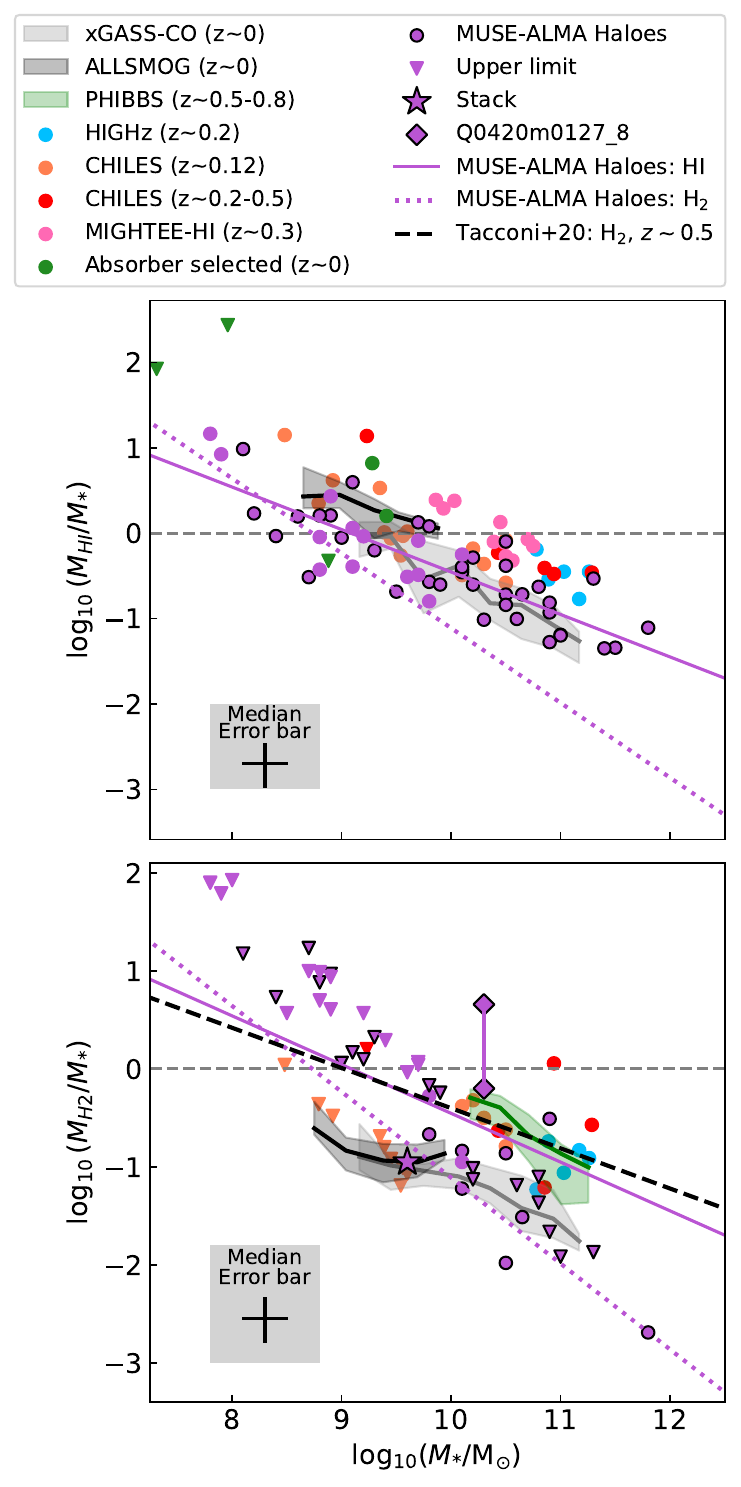}
    \caption{Atomic (upper panel) and molecular (lower panel) gas-to-stellar mass ratios versus stellar mass. {Black outlines indicate the MUSE-ALMA Haloes galaxies with the most robust stellar mass estimates.} {Linear fits (in log-log space) are shown for the MUSE-ALMA Haloes sample atomic gas fraction and molecular gas fraction by solid and dotted lines respectively.} Literature data sets outlined in Sect. \ref{sec:baryon_res} are shown for comparison:  {xGASS-CO \citep{Saintonge:17,Catinella:18}, ALLSMOG \citep{Cicone:17}, and PHIBSS \citep{Freundlich:19} are shown by their rolling median and interquartile range,  {the \cite{Tacconi:20} molecular gas fraction scaling relation is shown by the dashed line,} and absorber-selected $z\sim0$ \citep{Kulkarni:22}, CHILES \citep{Fernandez:16,Hess:25,Bluebird:26}, MIGHTEE-\HI\ \citep{Jarvis:25} and HIGHz \citep{Cortese:17} data are shown as individual points.}  }
       
    \vspace{-5pt}
    \label{fig:fgas}
\end{figure}

The atomic gas fractions of the MUSE-ALMA Haloes galaxies are consistent with the low to intermediate redshift comparison samples (xGASS, ALLSMOG, CHILES, HIGHz, MIGHTEE-\HI) within overlapping stellar mass range. {The best-fitting MUSE–ALMA Haloes \HI\ gas fraction relation aligns well with the rolling median of the volume-limited xGASS survey, which is constructed to be representative of the local galaxy population, while the ALLSMOG galaxies lie slightly above this relation, consistent with their selection as nearby star-forming galaxies.}  The $z>0$ \HI\ detections from CHILES, HIGHz  {and MIGHTEE-\HI} {also} predominantly lie above this relation. Given the sensitivity required to detect \HI\ emission at these redshifts, this offset is expected, as these surveys are naturally biased towards gas-rich systems. The agreement between the MUSE–ALMA Haloes relation and the $z\sim0$ samples likely reflects both the use of an \HI\ mass estimator calibrated at $z\sim0$ and the expectation that the cosmic \HI\ mass density does not evolve significantly over this redshift range \citep{Peroux:20,Walter:20}.

The molecular gas fractions of the MUSE–ALMA Haloes galaxies lie between those of the $z\sim0$ and $z\sim0.65$ comparison samples, but exhibit a steeper slope than the rolling medians of both. They also show systematically lower molecular gas fractions as well as a steeper slope (by $\sim0.4$) than the $z\sim0.5$ scaling relation of \cite{Tacconi:20}, derived from a compilation of literature data.

At the high-mass end, several sources from the MUSE-ALMA Haloes sample remain undetected in CO despite stringent upper limits that are comparable to or even lower than those reported in the literature, potentially indicating the presence of CO-dark molecular gas, as discussed by \cite{Bollo:26}. 

At low stellar masses, the MUSE–ALMA Haloes galaxies are expected to be gas-dominated, as is typically the case for low-mass systems at $z=0$ \citep{Catinella:18}.  {In this regime ($\log(M_*)<9.5$), the current $M_{\rm H2}$ upper limits probe a regime where CO emission is expected to be faint. Even low mass \HI-detected galaxies from CHILES are not detected in CO in this mass range, despite their  {stringent} $M_{\rm H2}$ limits. While CHILES adopts a Galactic $\alpha_{CO}$, applying a metallicity-dependent $\alpha_{CO}$ consistent with the MUSE–ALMA Haloes analysis would increase the molecular gas fraction upper limits by at most $\sim$0.3 dex, without altering this overall picture. Deeper observations in this regime would further constrain the molecular gas content of \HI\ absorption selected galaxies and could lead to a flatter inferred slope.} 

An alternative representation of the gas fraction using $f_{gas} = M_{\rm gas}/(M_{\rm gas}+M_{*})$, is presented in Appendix \ref{appen:gf}. While the MUSE-ALMA Haloes galaxies are most consistent with the $z\sim0$ scaling relation \citep{Tacconi:20,ForsterScheiber:20} if $M_{\rm gas}=M_{\rm H2}$, including the \HI\ component makes them appear significantly more gas rich, bringing them closer to the $z\sim1$ relation and further highlighting the importance of atomic gas.

\subsection{Baryon and dark matter fractions}

% describe how we calc fb
The baryon and dark matter fractions are determined by computing the total dynamical mass within $2R_e$ and evaluating the relative contribution of baryonic components within the same radius.  The dynamical mass is calculated as described in Sect. \ref{sec:MD}, where the circular velocity is derived from ionised gas emission lines that trace only the inner regions of the galaxies. We include the total stellar mass in the baryon budget, as the contribution from stars beyond $2R_e$ is minimal. For example, in a galaxy with a Sérsic index of 1, approximately $86\%$ of the flux is enclosed within this radius. The total molecular hydrogen mass is also included in the baryonic mass, as this has also been shown to reside within the inner regions of the stellar disc \citep[e.g.][]{Leroy:09,Bolatto:17,Hess:25}.  

 {We do not include \HI\ in the inner baryon mass budget.} {While \HI\ makes up a significant baryon component, it typically resides in an extended reservoir beyond the visible disc, as shown in Sect. \ref{sec:baryon_res}. As a result, its contribution within the inner regions probed by our dynamical measurements is expected to be small relative to the existing uncertainties.} \HI\ has been shown to saturate at $\sim 10$ M$_{\odot}$ pc$^2$, \citep{Bigiel:12} at the transition threshold between \HI\ and H$_2$.  {Assuming a radially flat \HI\ distribution at this density, the median \HI\ contribution is only $\sim 5\%$ of the total dynamical mass within $2R_e$, and is therefore negligible compared to the uncertainties in the baryon fraction. However, we note that this contribution becomes more significant in lower-mass systems: for galaxies with $M_{\mathrm{Dyn}} < 10^{10},M_{\odot}$, the median \HI\ fraction rises to $\sim 27\%$.

\begin{figure}[h!]
    \centering
    \begin{subfigure}[b]{0.49\textwidth}
        \centering
        \includegraphics[width=\linewidth]{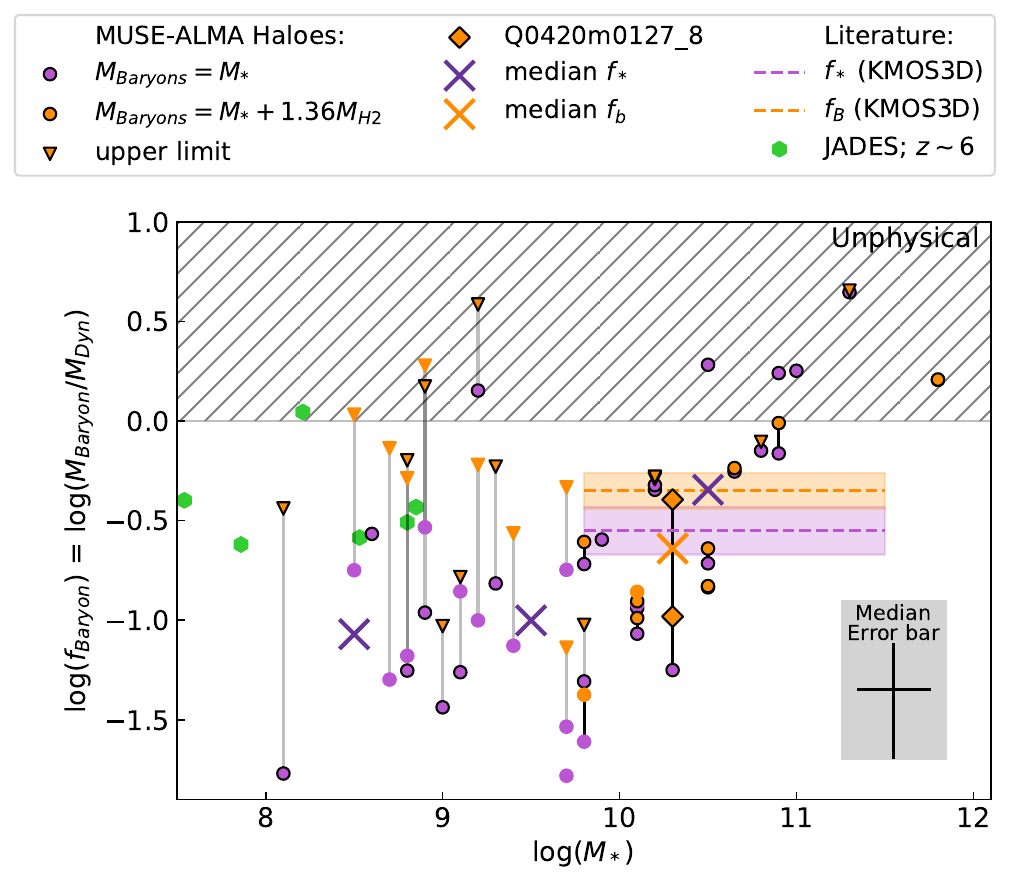}
        \caption{ }
        \label{fig:fbary_Mstar}
    \end{subfigure}%
    
    \begin{subfigure}[b]{0.49\textwidth}
        \centering
        \includegraphics[width=\linewidth]{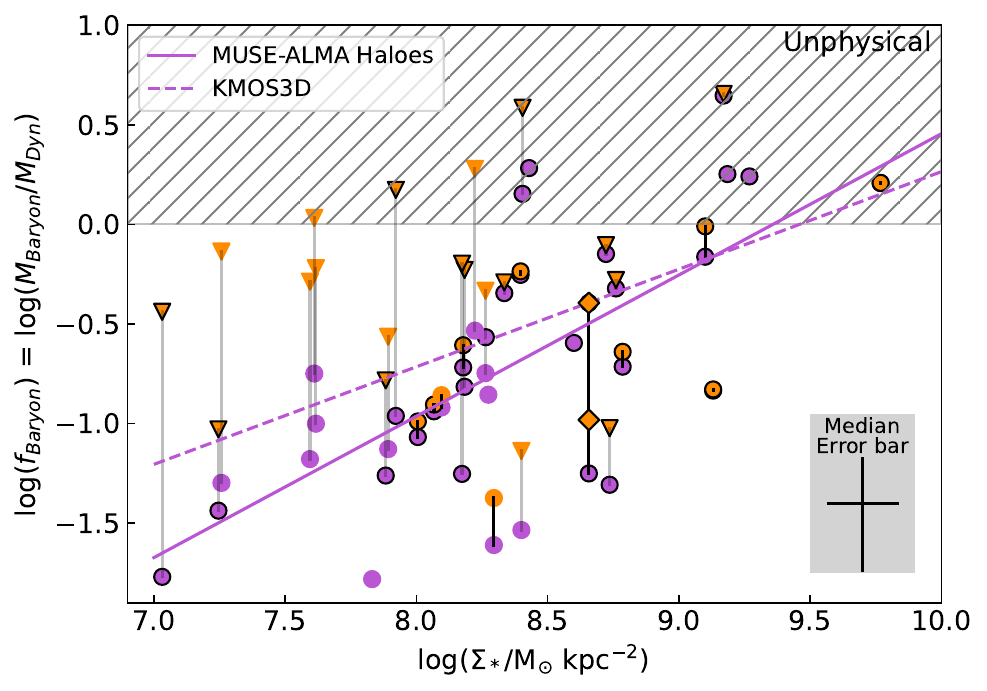}
        \caption{ }
        \label{fig:fbary_sigma}
    \end{subfigure}
    
    \caption{Baryon fraction (baryonic to dynamical mass ratio) versus stellar mass (upper panel) and stellar surface density (lower panel). {Baryon fractions are measured within 2$R_e$ for MUSE–ALMA Haloes and $R_e$ for literature samples.} Vertical lines show the change when including molecular gas (orange) compared to stellar mass only (purple). {Black outlines indicate the MUSE-ALMA Haloes galaxies with the most robust stellar mass estimates.} {The unphysical parameter space, where the baryon mass exceeds the total mass, is marked by the grey hashed regions. } {In the upper panel, the orange and purple dashed lines show the average baryon and stellar ratios of the $0.6<z<1.1$ KMOS$^{3D}$ sample \citep{Wuyts:16}.  {`x' markers indicate medians: the median for all CO-detected sources (baryon fraction), and medians in 1 dex stellar mass bins for the full sample (stellar mass fraction).}  In the lower panel, the solid and dashed lines show the best-fitting relation between stellar mass fraction and stellar mass surface density for the MUSE–ALMA Haloes and KMOS$^{3D}$ samples, respectively.  }}
    \vspace{-5pt}
    \label{fig:fb}
\end{figure}

{Figure \ref{fig:fb} shows the baryon fraction  {of the MUSE-ALMA Haloes sample} within $2R_e$ as a function of stellar mass  (a) and stellar mass surface density  (b).    Figure \ref{fig:fd_z} presents the dark matter fraction ($1-$baryon fraction) as function of redshift, adapted from \cite{deGraaff:24}. {In addition to observational measurements, this figure illustrates the evolution of high-redshift JWST analogues and MUSE-ALMA Haloes analogues selected from the TNG50 simulation \citep{Nelson:19,Pillepich:19}.} In Figs. \ref{fig:fb} and \ref{fig:fd_z}, literature observations shown for comparison adopt baryon fractions measured within $1R_e$, which systematically results in higher values compared to measurements within $2R_e$, although the offset is typically comparable to the uncertainties. This aperture dependence is also seen in the high-resolution kinematic models of the four $\log(M_*/\rm M_{\odot})\sim10.5$ MUSE-ALMA Haloes galaxies presented by \cite{Barfety:26}, for which the dark matter fraction increases by an average of 0.22 when measured within $2R_e$ rather than $1R_e$. Similarly, for the median dark matter fraction of the JWST descendants at $z=0.5$ ($\log(M_*/\rm M_{\odot})\sim10.6$) in TNG50, the dark matter fraction increases by 0.14 when calculated within $2R_e$ compared to $1R_e$.}

Uncertainties in the SED-derived stellar masses and adopted $\alpha_{\rm CO}$  affect the inferred baryon masses, while uncertainties in the kinematic and morphological modelling affect the dynamical masses. The MUSE-ALMA Haloes galaxies exhibit a large range of baryon fractions, as demonstrated by both {Figs.} \ref{fig:fb} and \ref{fig:fd_z}. In some cases, baryons account for almost all the mass within $2R_e$. Moreover, we find that six sources lie in an unphysical regime,  {with three of these remaining so beyond their error bars} (excluding upper limits). These sources represent $16\%$ {and $8\%$ respectively} of the sample shown here, and likely reflect the aforementioned uncertainties.  {At the opposite extreme}, several MUSE-ALMA Haloes sources are notably dark matter dominated. The most dark matter dominated source with a CO detection is Q1229m021$\_8$, in which baryons make up only $9\%$ of the total mass budget.  {Uncertainties in both the baryonic and dynamical masses remain important in this regime, particularly for low-mass systems, which are more susceptible to beam smearing effects in kinematic modelling and to metallicity-dependent variation in $\alpha_{\rm CO}$.}

% fb correlations - own results and comparison
An additional valuable comparison sample is the 240 star-forming disk galaxies  {($M_*\gtrsim10^{9.8}$)} at $0.6 < z < 2.6$ in the KMOS$^{3D}$ sample studied by \cite{Wuyts:16}. They use integral-field spectroscopic observations of H$\alpha$ emission to derive dynamical masses for their galaxies. While they do not have direct \HI\ or CO measurements, they infer the molecular gas mass from the depletion time, using an updated scaling relation from \cite{Genzel:15} which depends on redshift, SFR and stellar mass. 

We explore correlations between the inner baryon and dark matter fractions and various other galaxy properties. Figure \ref{fig:fbary_Mstar} shows the baryon fraction as a function of stellar mass. Low stellar mass galaxies are expected to be more dark matter dominated than their high stellar mass counterparts. Since they reside in shallower potential wells, they are less efficient at retaining condensed baryons due to processes such as supernova feedback and cosmic reionisation \citep{Papastergis:12,McQuinn:22}. This stellar mass dependence on the inner baryon fraction is seen in TNG50 and EAGLE \citep{deGraaff:24}. We find a moderate trend between stellar fraction and stellar mass in the MUSE-ALMA Haloes sample, with a Spearman rank coefficient of 0.6 (p$<1\times10^{-4}$), and a higher median baryon fraction in the highest stellar mass bin compared to lower-mass bins. {The Spearman rank correlation coefficient ranges from $-1$ to $+1$, where 0 indicates no monotonic correlation and $\pm1$ indicates a perfect monotonic relationship. The associated p-value is the probability of obtaining a correlation coefficient at least as extreme as the observed value from uncorrelated data, with smaller values indicating stronger evidence for a correlation.} \cite{Wuyts:16} divided the KMOS$^{3D}$ sample ($M_*>10^{9.8} \mathrm{M}_{\odot}$) into three redshift bins; their lowest-redshift bin ($0.6<z<1.1$) is most comparable to the MUSE-ALMA Haloes sample. They found a mean stellar mass fraction of $28\%$ and a mean baryon fraction of $45\%$, consistent with Illustris predictions. The median values for the MUSE-ALMA Haloes sample in the comparable stellar mass bin are consistent with these results. {The comparable baryon fractions despite being measured within a larger radius, as well as} the large scatter, particularly at low stellar masses, may partly reflect the \HI\ absorber-selected nature of our sample. This selection makes us sensitive to the gas-rich, low stellar mass systems that are often missed in optically-selected studies, increasing the number of high baryon fraction galaxies in the low stellar mass regime.  

 {We find that sources with high stellar surface densities  have the largest stellar and baryonic mass fractions, shown in Fig. \ref{fig:fbary_sigma}.  Similar behaviour has been reported in other samples over $0.3\lesssim z \lesssim 2.5$ \citep{Wuyts:16,Genzel:20,NestorShachar:23,Ciocan:26}. We measure a Spearmank rank coefficient of 0.7 (p$<1\times10^{-5}$), with a best fit relation of $f_*=(0.7\pm0.1)\log(\Sigma_*/$ M$_{\odot}$ kpc$^{-2}$) $- (6.6\pm1.0)$. This is broadly consistent with the results of KMOS$^{3D}$, although their slope is shallower by $\sim0.2$. Higher surface densities may be associated with flatter inner dark matter profiles driven by baryonic feedback, leading to higher baryon fractions \citep{NestorShachar:23}. This trend may also arise because the ionised gas kinematic tracer used to estimate the dynamical mass is probing the inner most baryon-dominated regions in compact sources (high surface densities). In contrast, the ionised gas probes further into the dark matter halo for galaxies with more extended stellar and baryonic distributions, resulting in lower baryon fractions \citep{Wuyts:16}.   }

\begin{figure}[h!]
    \centering
    \includegraphics[width=\linewidth]{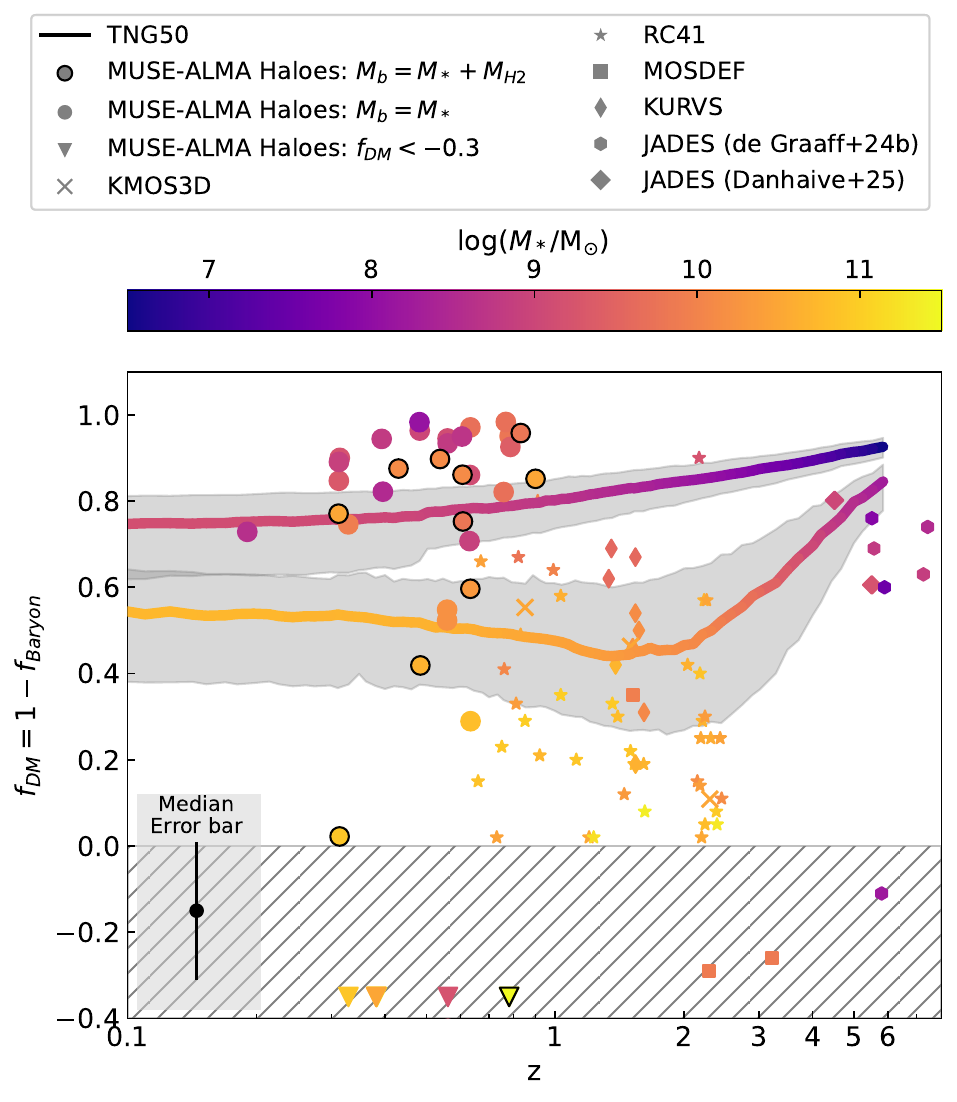}
    \caption{The dark matter fraction ($1-$baryon fraction) as function of redshift, adapted from \cite{deGraaff:24}. {Baryon fractions are measured within 2$R_e$ for MUSE–ALMA Haloes and $R_e$ for literature samples.} MUSE-ALMA Haloes sources that were detected in CO have a baryon mass that includes both stellar mass and molecular gas mass, and are shown by the markers with a black outline. For those not detected in CO, the stellar mass is assumed to dominate the baryon mass. Observational data of `typical' star-forming galaxies around cosmic noon are shown: { KMOS$^{3D}$ \citep[][]{Wuyts:16}, RC41 \citep[][]{Genzel:20}, MOSDEF \citep{Price:20}, and KURVS \citep{Puglisi:23} }  and at higher redshift:  {JADES} \citep{degraaff:24a,Danhaive:25c}. Median values are shown for the KMOS$^{3D}$ and JADES \citep{Danhaive:25c} samples {within several redshift bins}. {The redshift evolution of the dark matter fraction within $2R_e$ for galaxy analogues in TNG50 is shown by the solid lines, coloured by stellar mass. The lower line shows analogues of dark matter–dominated low-mass galaxies observed with JWST at $z\sim6$ evolved to $z=0$, while the upper line shows analogues of the MUSE-ALMA Haloes galaxies selected at $z\sim0.5$ with $8.75 < \log(M_\star/M_\odot) < 9.25$.  The grey shaded regions show the 16th–84th percentiles.} {The unphysical parameter space, where the baryon mass exceeds the total mass, is marked by the grey hashed regions. }    }
    \vspace{-5pt}
    \label{fig:fd_z}
\end{figure}

 {We place the baryon content of the MUSE-ALMA Haloes galaxies in the context of redshift evolution in Fig. \ref{fig:fd_z}.  \cite{deGraaff:24} study the evolution of low-mass ($\log(M_*/\rm M_{\odot})=8-9$) galaxies from $z=6$ to $z=0$ using the TNG50 simulation, selected as analogues of the dark matter-dominated low-mass galaxies observed with JADES/JWST at $z\sim6$ \citep{degraaff:24a}. {In Fig. \ref{fig:fd_z}} the median evolutionary track of the dark matter fraction is shown by the solid line, coloured by the stellar mass. While \cite{degraaff:24a} present the TNG evolution for measurements within 1 $R_{\rm e}$, here we recalculate the dark matter fraction within 2 $R_{\rm e}$ for 434 simulated JADES analogues to enable a direct comparison with the MUSE-ALMA Haloes sample. We additionally show the median evolutionary track of 1461 TNG analogues selected to match the MUSE-ALMA Haloes galaxies at $z\sim0.5$ with $8.75 < \log(M_\star/M_\odot) < 9.25$, following both their main progenitor branches and descendants. } 
 
 {The MUSE-ALMA Haloes sample helps fills the previously underexplored parameter space at low to intermediate redshift. The TNG50 evolutionary track suggests that the low-mass galaxies detected with JWST in JADES at $z\sim4-6$ \citep{degraaff:24a,Danhaive:25c} may represent progenitors of higher mass centrally baryon-dominated systems at cosmic noon \citep{deGraaff:24}. For comparison, we show several observational data sets at around cosmic noon, including the medians from the KMOS$^{3D}$ sample from \cite{Wuyts:16}, the RC41 sample, \citep{Genzel:20}, the MOSDEF survey \citep{Price:20}, and the KURVS sample \citep{Puglisi:23}. Some of these systems, as well as three of the MUSE–ALMA Haloes galaxies, are broadly consistent with the TNG50 predictions for the descendants of the JWST-like galaxies. However, most of the MUSE–ALMA Haloes systems appear to be more strongly dark-matter dominated.  {Many of the MUSE–ALMA Haloes galaxies with high dark matter fractions have lower stellar masses than those in other intermediate-redshift surveys, making them more comparable to the high-redshift, dark matter-dominated galaxies observed in JADES. More MUSE-ALMA Haloes galaxies are consistent with the lower-mass TNG50 track. The progenitors of these analogues are generally too low mass to be detectable in current JWST observations. Nevertheless, approximately half of the MUSE-ALMA Haloes sample remains more dark matter dominated than the TNG50 analogues.}  We do not observe {a} clear redshift evolution in the dark matter fractions of the MUSE–ALMA Haloes galaxies, although the limited sample size prevents us from controlling for additional variables such as stellar mass surface density. These results are laying the groundwork for future high-resolution observations of larger samples to further clarify the evolution of dark-matter fraction over cosmic time. }

 {Reconciling low- and high-redshift observations, as well as observations with simulations, remains a key challenge. Semi-analytical models, such as SHARK \citep{Lagos:18}, provide empirically motivated prescriptions for galaxy evolution. Many large-volume simulations do not explicitly resolve cold gas (T $< 10^4$ K), instead relying on post-processing prescriptions, which can result in tensions with observations \citep[e.g.][]{Diemer:19,Popping:19}. This is now beginning to change, with new simulations such as COLIBRE  \citep{Schaye:25,Lagos:25} and \textsc{ColdSIM} \citep{Maio:22}, which include a more explicit treatment of cold gas, such as non-equilibrium chemistry. Observational datasets such as the MUSE–ALMA Haloes survey are therefore critical for constraining baryon budgets and benchmarking these simulation predictions. }

%%%%%%%%%%%%%%%%%%%%%%%%%%%%%%%%%%%%%%%%%%%%%%%%%%%%%%%
\section{Summary and conclusions}
\label{sec:concl}

In this work, we characterise the baryon content of galaxies' ISM and CGM in the MUSE-ALMA Haloes survey.   This survey  {is designed to connect galaxies to their surrounding gas by combining MUSE, ALMA and HST observations.}  {The properties of the \HI\ absorbers, together with their frequent association with multiple galaxies, provide key constraints on the CGM and local environment, which we directly connect to the ISM gas properties traced by our ionised and molecular gas observations.} This paper brings together the wealth of data collected on this galaxy sample over the past several years to characterise the baryonic content and gas phases. We derive new quantities, including estimated \HI\ and dynamical masses, and revised group masses.

Our key results are as follows:
\vspace{-5pt}
\begin{enumerate}
    \item  We estimate the \HI\ masses following the prescription from \cite{Zu:20}, based on the $g-r$ colour and stellar mass. For sources with impact parameters $\lesssim100$ kpc there is a correlation with the \HI\ masses inferred from absorber properties, following \cite{Wang:26},  albeit with large scatter that is influenced by the local overdensity environment (Fig. \ref{fig:MHI_MHI}).   {Both approaches provide valuable, independent means of estimating \HI\ masses for intermediate-redshift galaxies.} 

    \item The MUSE–ALMA Haloes galaxies are expected to host larger reservoirs of \HI\ than H$_2$, or comparable amounts of the two phases (Fig. \ref{fig:MHIMH2}).  {Thus, \HI\ makes up a significant fraction of the condensed baryonic content of galaxies,  {especially when considering the full extent of ISM beyond $\sim 2R_e$ and into the CGM,}  {and remains an important component in} studies at intermediate redshift.}

    \item { We investigate the effect of overdensity mass on the properties of individual galaxies, however we find no correlation between the group masses. Given the limited dynamic range, together with the heterogeneous nature of our sample, the absence of clear trends is expected.}

    \item   {The MUSE–ALMA Haloes galaxies span four orders of magnitude in stellar mass and exhibit a clear decline in both atomic and molecular gas fractions with increasing stellar mass (Fig. \ref{fig:fgas}). This suggests that stellar mass builds up more rapidly than the cold gas reservoir.  The linear fits indicate that the H$_2$ gas fraction declines more steeply with stellar mass than the \HI\ gas fraction,  consistent with massive galaxies consuming their molecular gas more efficiently or with an inefficient conversion of \HI\ to H$_2$. Overall both relations are broadly consistent with low- to intermediate-redshift comparison samples.}

    \item   {We find a moderate trend between baryon fraction and stellar mass in the MUSE-ALMA Haloes sample (Fig. \ref{fig:fbary_Mstar}), with a Spearman rank coefficient of 0.6 (p$<1\times10^{-4}$), and a higher median baryon fraction in the highest stellar mass bin compared to lower-mass bins. The large scatter, particularly at low stellar masses, are likely driven by the uncertainties discussed and the \HI\ absorber-selection of the sample, which includes gas-rich low-mass galaxies often missed in optical surveys. }

    \item  {We find that sources with high stellar surface densities have the largest stellar and baryonic mass fractions (with a Spearman rank coefficient of 0.7, p$<1\times10^{-5}$), shown in Fig. \ref{fig:fbary_sigma}.  This trend may reflect the impact of baryonic feedback producing flatter inner dark matter profiles in high surface-density systems, or aperture effects, where the ionised gas kinematic tracer used to estimate the dynamical mass is probing the innermost baryon-dominated regions in compact sources. }

    \item  The MUSE–ALMA Haloes sample fills a gap in baryon fraction measurements at low-to-intermediate redshift, providing important context for evolutionary studies (Fig. \ref{fig:fd_z}). A subset of the sample is consistent with TNG50 predictions for evolved JWST-like $z\sim6$ low-mass galaxies, while many systems appear more strongly dark matter dominated with lower masses, whose progenitors would be undetectable with JWST. 
    
\end{enumerate}

The MUSE-ALMA Haloes programme demonstrates the  power of combining multiphase gas tracers, neutral, molecular, and ionised gas, within an absorber-selected sample, enabling a more complete view of the baryon cycle in galaxies than is typically possible.  Upcoming intermediate-redshift absorber surveys, such as ByCycle/4MOST \citep{Peroux:23}, will deliver substantially larger samples and unprecedented statistical power. This highlights the importance of detailed, multi-wavelength studies like this one in preparing us to better interpret and understand these populations.  {In this context, our results advance both the baryon census and our understanding of the baryon cycle in galaxies by constraining the gas within the ISM and {relating it to the circumgalactic gas traced by absorption}.} Looking ahead, the Square Kilometre Array will extend direct \HI\ measurements to higher redshifts, complimenting this work in understanding the role of neutral gas in galaxy evolution across cosmic time  \citep[][]{Braun:15}. Together, these advances will place absorber-selected galaxies in a critical position for connecting small-scale gas physics with the broader picture of galaxy evolution.

\begin{acknowledgements}
The authors gratefully thank Anna de Graaff for sharing the data used in Fig. \ref{fig:fd_z}. 

C.B. acknowledges funding by the European Union (ERC Advanced Grant GALPHYS, 101055023). Views and opinions expressed are, however, those of the author(s) only and do not necessarily reflect those of the European Union or the European Research Council. Neither the European Union nor the granting authority can be held responsible for them. RA acknowledges funding from the European Research Council (ERC) under the European Union's Horizon 2020 research and innovation programme (grant agreement 101020943, SPECMAP-CGM). 
\end{acknowledgements}

   \bibliographystyle{aa} % style aa.bst
   \bibliography{aanda} % your references Yourfile.bib

\begin{appendix}

\section{Galaxy properties}
\label{appen:mega_plot}

Figure \ref{fig:mega_plot} summarises the properties available for each galaxy in the MUSE-ALMA Haloes sample. Due to the evolving nature of the programme, as well as varying sensitivities and resolutions, all properties are available for some of 79 galaxies. Table \ref{tab:MHIMD} presents the \HI\ and dynamical masses derived for this study. Descriptions of how each galaxy property is derived is outlined in Sect. \ref{sec:quantities}. 

\begin{sidewaysfigure*}
    \centering
    \includegraphics[width=\textheight]{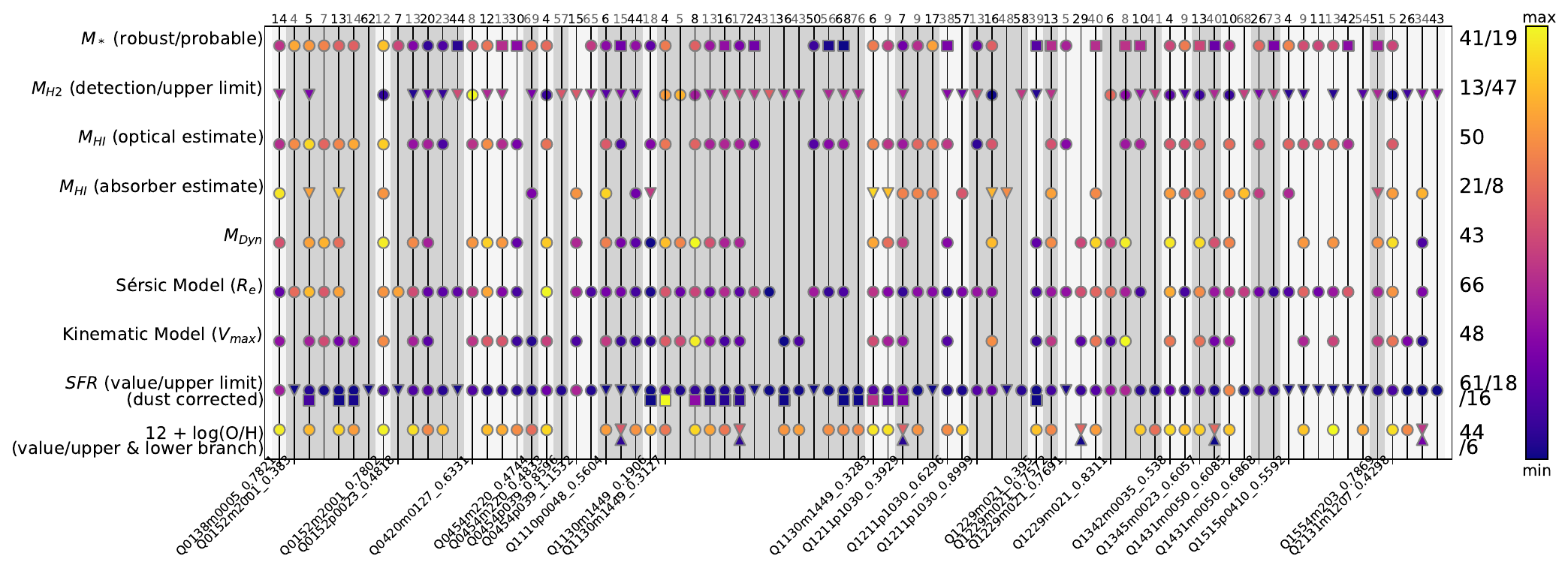}  % width set to page height
    \caption{Summary of the available properties for each galaxy in the MUSE–ALMA Haloes sample. Each vertical line represents a different galaxy, grouped by absorber and alternating between grey and white shading. Absorber systems are indicated along the bottom in the format Quasar$\_$absorber–redshift, and the corresponding galaxy ID numbers are listed across the top. From top to bottom, the galaxy properties shown are stellar mass, molecular hydrogen mass, {\HI\ mass estimated from optical properties, \HI\ mass estimated from absorber properties (for sources with $b<100$ kpc and $N_{HI}<10^{20.5}$ cm$^{-2}$)} , dynamical mass, Sérsic model, kinematic model, star formation rate, $12+\log(\rm O/H)$ metallicity. The numbers listed on the right-hand side indicate the total number of galaxies for which each property is available. Points are coloured by the normalised value of each quantity. Upper limits for the $M_{\rm H2}$ and SFR are shown with inverted triangle markers.  Probable stellar masses and dust-corrected SFRs are indicated by square markers. Metallicities with unresolved upper–lower branch degeneracies are shown by triangle markers. }
    \label{fig:mega_plot}
\end{sidewaysfigure*}

\begin{table*}[t]
\centering
\caption{Galaxy properties derived in this study. From left to right: (1) The galaxy ID. (2) The \HI\ mass derived from optical properties \citep[$g-r$ and stellar mass;][]{Zu:20}. (3) The \HI\ mass derived from absorber properties \citep[impact parameter and \HI\ column density;][]{Wang:26}. Only values for systems with impact parameter $<100$ kpc and \HI\ column density $<10^{20.5}$ cm$^{-2}$ are provided. (4) The dynamical mass within $2R_e$.}
\label{tab:MHIMD}
\setlength{\tabcolsep}{2pt}
\begin{minipage}{0.495\linewidth}
\centering
\begin{tabular}{lccc}
\hline
Galaxy ID & $\log(\frac{M_{\rm HI~optical}}{\rm M_{\odot}})$ &$\log(\frac{M_{\rm HI~absorber}}{\rm M_{\odot}})$ & $\log(\frac{M_{\rm Dyn~<2Re}}{\rm M_{\odot}})$  \\
\hline
Q0138m0005$\_$14 & $9.2\pm 0.1$ & $10.9\pm 0.1$ & $11.1\pm0.2$ \\
Q0152m2001$\_$4 & $10.2\pm 0.1$ & ... & ... \\
Q0152m2001$\_$5 & $10.8\pm 0.1$ & $<10.4\pm 0.1$ & $10.7\pm0.2$ \\
Q0152m2001$\_$7 & $9.8\pm 0.1$ & ... & $10.7\pm0.1$ \\
Q0152m2001$\_$13 & $10.1\pm 0.1$ & $<10.7\pm 0.1$ & $10.2\pm0.3$ \\
Q0152m2001$\_$14 & $10.4\pm 0.1$ & ... & ... \\
Q0152m2001$\_$62 & ... & ... & ... \\
Q0152m2001$\_$12 & $10.7\pm 0.1$ & $10.2\pm 0.1$ & $11.6\pm0.1$ \\
Q0152p0023$\_$7 & ... & ... & ... \\
Q0152p0023$\_$13 & $8.9\pm 0.1$ & ... & $10.4\pm0.2$ \\
Q0152p0023$\_$20 & $9.1\pm 0.1$ & ... & $9.9\pm0.3$ \\
Q0152p0023$\_$23 & $8.4\pm 0.1$ & ... & ... \\
Q0152p0023$\_$44 & ... & ... & ... \\
Q0420m0127$\_$8 & $9.3\pm 0.1$ & ... & $11.6\pm0.1$ \\
Q0420m0127$\_$12 & $10.2\pm 0.1$ & ... & $10.9\pm0.1$ \\
Q0420m0127$\_$13 & $9.2\pm 0.1$ & ... & $11.2\pm0.1$ \\
Q0420m0127$\_$30 & $8.7\pm 0.1$ & ... & $10.0\pm0.3$ \\
Q0454m220$\_$69 & ... & $8.7\pm 0.1$ & ... \\
Q0454m220$\_$4 & $9.9\pm 0.1$ & ... & $10.9\pm0.1$ \\
Q0454p039$\_$57 & ... & ... & ... \\
Q0454p039$\_$15 & ... & $10.2\pm 0.1$ & $11.2\pm0.1$ \\
Q0454p039$\_$65 & ... & ... & ... \\
Q1110p0048$\_$6 & $9.7\pm 0.1$ & $10.7\pm 0.1$ & $10.4\pm0.2$ \\
Q1110p0048$\_$15 & $8.4\pm 0.1$ & ... & $10.0\pm0.3$ \\
Q1110p0048$\_$44 & ... & $8.7\pm 0.1$ & $9.0\pm0.6$ \\
Q1130m1449$\_$18 & $8.8\pm 0.1$ & $<9.4\pm 0.1$ & $9.2\pm0.6$ \\
Q1130m1449$\_$4 & $10.0\pm 0.1$ & ... & $11.1\pm0.1$ \\
Q1130m1449$\_$5 & ... & ... & $10.7\pm0.2$ \\
Q1130m1449$\_$8 & $9.8\pm 0.1$ & ... & $11.2\pm0.1$ \\
Q1130m1449$\_$13 & $9.1\pm 0.1$ & ... & $10.1\pm0.2$ \\
Q1130m1449$\_$16 & $9.2\pm 0.1$ & ... & $10.2\pm0.2$ \\
Q1130m1449$\_$17 & $9.1\pm 0.1$ & ... & $9.9\pm0.3$ \\
Q1130m1449$\_$24 & $8.8\pm 0.1$ & ... & ... \\
Q1130m1449$\_$31 & ... & ... & ... \\
Q1130m1449$\_$36 & ... & ... & ... \\
Q1130m1449$\_$43 & ... & ... & ... \\
Q1130m1449$\_$50 & $8.4\pm 0.1$ & ... & ... \\
Q1130m1449$\_$56 & $8.8\pm 0.1$ & ... & ... \\
Q1130m1449$\_$68 & $9.0\pm 0.1$ & ... & ... \\
Q1130m1449$\_$76 & ... & ... & ... \\
\hline
\end{tabular}
\end{minipage}
%\hfill
\begin{minipage}{0.495\linewidth}
\centering
\begin{tabular}{lccc}
\hline
Galaxy ID & $\log(\frac{M_{\rm HI~optical}}{\rm M_{\odot}})$ &$\log(\frac{M_{\rm HI~absorber}}{\rm M_{\odot}})$ & $\log(\frac{M_{\rm Dyn~<2Re}}{\rm M_{\odot}})$ \\
\hline
Q1130m1449$\_$6 & $10.1\pm 0.1$ & $<10.8\pm 0.1$ & $10.7\pm0.1$ \\
Q1130m1449$\_$9 & $9.3\pm 0.1$ & $<10.6\pm 0.1$ & $10.5\pm0.2$ \\
Q1211p1030$\_$7 & $9.0\pm 0.1$ & $10.1\pm 0.1$ & $10.1\pm0.3$ \\
Q1211p1030$\_$9 & $9.8\pm 0.1$ & $10.1\pm 0.1$ & ... \\
Q1211p1030$\_$17 & $10.1\pm 0.1$ & $10.1\pm 0.1$ & ... \\
Q1211p1030$\_$38 & $9.3\pm 0.1$ & ... & $9.4\pm0.4$ \\
Q1211p1030$\_$57 & ... & $9.7\pm 0.1$ & ... \\
Q1211p1030$\_$13 & $8.2\pm 0.1$ & ... & ... \\
Q1211p1030$\_$16 & $9.7\pm 0.1$ & $<10.5\pm 0.1$ & $11.3\pm0.1$ \\
Q1211p1030$\_$48 & ... & $<10.2\pm 0.1$ & ... \\
Q1211p1030$\_$58 & ... & ... & ... \\
Q1229m021$\_$39 & ... & ... & $9.2\pm0.5$ \\
Q1229m021$\_$13 & $9.6\pm 0.1$ & $10.4\pm 0.1$ & $10.4\pm0.2$ \\
Q1229m021$\_$5 & $8.8\pm 0.1$ & ... & ... \\
Q1229m021$\_$29 & ... & ... & $10.2\pm0.2$ \\
Q1229m021$\_$40 & ... & $10.2\pm 0.1$ & $11.5\pm0.1$ \\
Q1229m021$\_$6 & ... & ... & $11.7\pm0.1$ \\
Q1229m021$\_$8 & $9.0\pm 0.1$ & ... & $11.4\pm0.1$ \\
Q1229m021$\_$10 & $9.1\pm 0.1$ & ... & ... \\
Q1229m021$\_$41 & ... & ... & ... \\
Q1342m0035$\_$4 & $9.7\pm 0.1$ & $10.3\pm 0.1$ & $11.2\pm0.1$ \\
Q1342m0035$\_$9 & $9.6\pm 0.1$ & $9.8\pm 0.1$ & ... \\
Q1345m0023$\_$13 & $9.9\pm 0.1$ & $10.3\pm 0.1$ & $11.0\pm0.1$ \\
Q1345m0023$\_$40 & ... & ... & $10.0\pm0.3$ \\
Q1431m0050$\_$10 & $9.9\pm 0.1$ & $10.2\pm 0.1$ & $10.5\pm0.2$ \\
Q1431m0050$\_$68 & ... & $10.6\pm 0.1$ & ... \\
Q1431m0050$\_$26 & $9.6\pm 0.1$ & $9.4\pm 0.1$ & ... \\
Q1431m0050$\_$73 & ... & ... & ... \\
Q1515p0410$\_$4 & $9.8\pm 0.1$ & $9.2\pm 0.1$ & ... \\
Q1515p0410$\_$9 & $9.6\pm 0.1$ & ... & $10.5\pm0.2$ \\
Q1515p0410$\_$11 & $9.6\pm 0.1$ & ... & ... \\
Q1515p0410$\_$13 & $9.9\pm 0.1$ & ... & $10.5\pm0.1$ \\
Q1515p0410$\_$42 & $9.2\pm 0.1$ & ... & ... \\
Q1515p0410$\_$54 & ... & ... & ... \\
Q1554m203$\_$51 & ... & $<9.6\pm 0.1$ & $10.5\pm0.2$ \\
Q2131m1207$\_$5 & $9.7\pm 0.1$ & $10.3\pm 0.1$ & $11.0\pm0.1$ \\
Q2131m1207$\_$26 & ... & ... & ... \\
Q2131m1207$\_$34 & ... & $10.5\pm 0.1$ & $9.5\pm0.4$ \\
Q2131m1207$\_$43 & ... & ... & ... \\
&&&\\
\hline
\end{tabular}
\end{minipage}
\end{table*}

\section{Estimating \HI\ mass}
\label{appen:HI}

 {The \HI\ content of a galaxy is linked to its star-forming properties through the conversion of gas from the extended \HI\ reservoir into molecular gas.} As such, the \HI\ mass (and the \HI\ to stellar mass ratio), has been shown to correlate with a number of different galaxy properties, albeit with some scatter, including individual estimators or linear combinations of optical, infrared and UV colours, surface brightness, absolute magnitude and size \citep[e.g.][]{Kannappan:04,Catinella:10,Catinella:13,Toribio:11,Li:12,Denes:14,Eckert:15,Zu:20,Zhang:09,Zhang:21}. As well as empirical calibrations, machine learning algorithms have also been used to estimate the \HI\ content of galaxies from these properties \citep{Teimoorinia:17,Rafieferantsoa:18}.  Previous studies have demonstrated that the NUV$-r$ colour is one of the properties most strongly correlated with the \HI\ to stellar mass ratio \citep{Catinella:10,Catinella:13}, and these relationships, calibrated at $z=0$, have been applied to estimate \HI\ masses in high redshift samples \citep[e.g][]{Zhang:21}. Unfortunately, deep UV data are not available for our sample.  The Galaxy Evolution Explorer \citep[GALEX;][]{Martin:05} provides imaging in the near ultraviolet (NUV; 1770-2730 $\mathring{A}$) and the far ultraviolet (FUV; 1350-1780 $\mathring{A}$), however, the MUSE-ALMA Haloes galaxies are not detected (to a depth of $\sim20.5$ mag).  Instead, we adopt the \HI\ to stellar mass ratio as a function $g-r$ colour and stellar mass, following \cite{Zu:20}, which provides a well-calibrated and empirically robust alternative for estimating \HI\ content. 

In our investigation of the most appropriate \HI\ mass estimator for the MUSE–ALMA Haloes galaxies, we compared the \HI\ masses derived from the \cite{Zu:20} relation with those estimated by the relation from \citet{Denes:14}. In contrast to \cite{Zu:20}, who use $g - r$ colour and stellar mass, \cite{Denes:14} explore \HI\ mass scaling relations with $B, R, I, J, H$ and $K$-band magnitudes and $B, R, I, J$ and $K$-band diameters with galaxies from HIPASS. Among the magnitude relations, the $B$-band relation has the lowest scatter ($\sigma=0.26$), and is given {by} this equation: $\log(M_{\rm HI}) = 2.89 - 0.34 M_B$.

To minimise environmental effects, \cite{Denes:14} restrict their analysis to galaxies in low-density environments ({defined by a projected 7th nearest-neighbour surface density of} $\Sigma_7 < 1$ Mpc$^{-2}$). They also test the impact of using a volume-limited sample and find that the resulting fits are in good agreement with those from their non-volume-limited sample. However, they do not adopt a volume-limited sample because the number of galaxies becomes too small.  Furthermore, \cite{Chowdhury:24} compare results of \HI\ stacking at $z=1$ to this relation. They find that {their} results are consistent within $2\sigma$ significance (see Fig. 3 in their paper). 

However, because this relation is based on a single magnitude and calibrated using predominantly blue HIPASS galaxies, it tends to overestimate the \HI\ masses of red galaxies. Figure \ref{fig:MHI_compare} compares the \HI\ masses estimated by the \cite{Zu:20} method, which we adopt for our study, and those estimated using the \cite{Denes:14} method.  While the majority of blue galaxies have estimated \HI\ masses that agree within 0.3 dex (roughly a factor of $\sim2$, indicated by the shaded region), a few galaxies, particularly those with the largest stellar masses, show higher \HI\ masses when estimated using the \cite{Zu:20} method. In contrast, several red galaxies exhibit \HI\ masses that are significantly overestimated by the \cite{Denes:14} method. Based on this analysis, we conclude that the \cite{Zu:20} prescription provides the most reliable estimations for the \HI\ masses of the MUSE-ALMA Haloes sample.

\begin{figure}[h!]
    \centering
    \includegraphics[width=\linewidth]{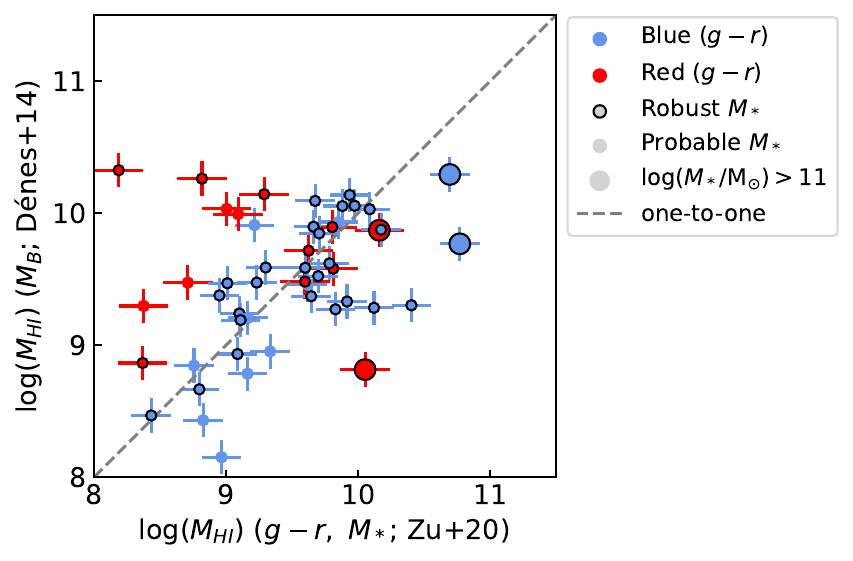}
    \caption{ {Comparison} of the \HI\ masses estimated by the \cite{Zu:20} method (using $g-r$ colour and $M_*$), which we adopt for our study, and those estimated using the \cite{Denes:14} method (using $M_B$). Galaxies are coloured according to the stellar mass dependent blue/red cut defined in Sect. \ref{sec:MHI}, and large markers indicate galaxies with stellar masses $>10^{11}$ M$_{\odot}$.  }
    \label{fig:MHI_compare}
\end{figure}

\section{Absorber systems}
\label{appen:groups}

Table \ref{tab:groups} summarises the properties of each absorber system (Sect. \ref{sec:groupmass}). It lists the number of galaxies associated to the absorber system in the MUSE primary sample (within the 1 arcmin MUSE field of view and 500 km s$^{-1}$ of the absorber), as well as the number of galaxies in the overdensity in the extended field ({within the MUSE field of view and 1000 km s$^{-1}$ or} within the HST field of view and 1000 km s$^{-1}$).  The table also includes the corresponding group velocity dispersions and the inferred group masses for both scenarios.

{The group velocity dispersions were calculated using the \textsc{gapper} method, as described in Sect. \ref{sec:MD}. Although the \textsc{gapper} estimator is optimised for groups with $\sim10$ members \citep{Beers:90}, it has been shown to provide useful velocity dispersion estimates for smaller systems \citep[e.g.][]{Yang:07,Cucciati:10,Robotham:11}. Nevertheless, the uncertainty in the derived velocity dispersions increases as the number of group members decreases. Additionally, as discussed in Sect. \ref{sec:MD}, our catalogue identifies galaxy overdensities around absorbers rather than uniformly defined, virialised groups. }

{We investigate the impact of the maximum velocity cut using both galaxies with spectroscopic redshifts from MUSE and galaxies with photometric redshifts from HST. We note that the typical uncertainty of the photometric redshifts is larger than the $\pm1000$ km s$^{-1}$ velocity window used to calculate the group mass upper limits (the typical uncertainty of $\Delta z\sim0.02$ corresponds to $\sim4000$ km s$^{-1}$ at $z\sim0.5$). We therefore vary the maximum velocity separation, $v_{\rm max}$, from the 500 km s$^{-1}$ adopted for the primary sample up to 4000 km s$^{-1}$. Figure \ref{fig:dv_hist} shows the distribution of velocity separations between galaxies and absorbers across all systems, distinguishing the primary MUSE sources, MUSE sources with larger velocity separations, and HST sources. The distribution shows that the vast majority of associated galaxies lie within 500 km s$^{-1}$ of the absorber. Figure \ref{fig:mgroup} shows how the derived group mass varies with $v_{\rm max}$, considering only spectroscopic redshift MUSE sources in panel (a) and both spectroscopic redshift MUSE and photometric redshift HST sources in panel (b). We find that for $v_{\rm max}\gtrsim1000$ km s$^{-1}$, a significant number of likely unbound galaxies are included, resulting in group masses substantially larger than expected for an absorption selected sample \citep[e.g.][]{Nielsen:18}. Therefore, while the derived velocity dispersions and group masses are subject to these limitations, they provide a useful measure of the relative environments of the absorber systems and can be used to investigate broad environmental trends. }

\begin{table*}[h!]

    \caption{Properties of galaxy groups linked to the absorber systems. From left to right the columns are as follows: (1) the Quasar field, (2) {$z_{\rm abs}$,} the redshift of the absorber, (3) $N_{\rm min}$, the number of galaxies associated to the absorber system in the MUSE primary sample (within the 1 arcmin MUSE field of view and 500 km s$^{-1}$ of the absorber), (4) {$N_{\rm max}$,} the number of galaxies in the overdensity in the extended field {(within the MUSE field of view and 1000 km s$^{-1}$ or the HST field of view and 1000 km s$^{-1}$)}, (5) {$\sigma_{\rm group,~min}$,} the group velocity dispersion corresponding to $N_{\rm min}$, (6) {$\sigma_{\rm group,~max}$,} the group velocity dispersion corresponding to $N_{\rm max}$ {(including galaxies with photometric redshifts within the HST field of view {and galaxies with spectroscopic redshifts within the MUSE field of view within $\pm 1000$ km s$^{-1}$})}, (7) {$M_{\rm group,~min}$,} the group mass corresponding to $N_{\rm min}$, and (8) {$M_{\rm group,~max}$,}  the group mass corresponding to $N_{\rm max}$ {(using $\sigma_{\rm group,~max}$ and the radius inferred from the location of all members)}. For overdensities with less than three members, the group velocity dispersion and mass are not calculated. }
    \label{tab:groups}
    
    \centering
    \begin{tabular}{cccccccc}
    \hline
        Quasar & $z_{\rm abs}$ & N$_{\rm min}$ & N$_{\rm max}$ & $\frac{\sigma_{\rm group, min}}{\rm km~s^{-1}}$ & $\frac{\sigma_{\rm group, max}}{\rm km~s^{-1}}$ & log($\frac{M_{\rm group, min}}{\rm M_{\odot}}$) & log($\frac{M_{\rm group, max}}{\rm M_{\odot}}$) \\
        \hline
         Q0152m2001 & 0.7802 & 1 & 1 & ... & ... & ... & ... \\ 
Q0454m220 & 0.4744 & 1 & 1 & ... & ... & ... & ... \\ 
Q0454m220 & 0.4833 & 1 & 1 & ... & ... & ... & ... \\ 
Q1130m1449 & 0.1906 & 1 & 1 & ... & ... & ... & ... \\ 
Q1229m021 & 0.395 & 1 & 1 & ... & ... & ... & ... \\ 
Q1229m021 & 0.7572 & 1 & 1 & ... & ... & ... & ... \\ 
Q1554m203 & 0.7869 & 1 & 1 & ... & ... & ... & ... \\ 
Q0138m0005 & 0.7821 & 1 & 2 & ... & ... & ... & ... \\ 
Q0454p039 & 0.8596 & 1 & 2 & ... & ... & ... & ... \\ 
Q0454p039 & 1.1532 & 2 & 2 & ... & ... & ... & ... \\ 
Q1130m1449 & 0.3283 & 2 & 2 & ... & ... & ... & ... \\ 
Q1211p1030 & 0.6296 & 2 & 3 & ... & 98.5 & ... & 11.9$\pm$0.3 \\ 
Q1431m0050 & 0.6085 & 2 & 3 & ... & 405.4 & ... & 13.1$\pm$0.1 \\ 
Q1345m0023 & 0.6057 & 2 & 4 & ... & 536.0 & ... & 13.8$\pm$0.1 \\ 
Q1431m0050 & 0.6868 & 2 & 4 & ... & 372.6 & ... & 13.5$\pm$0.1 \\ 
Q1342m0035 & 0.538 & 2 & 8 & ... & 145.4 & ... & 12.8$\pm$0.2 \\ 
Q1110p0048 & 0.5604 & 3 & 3 & 155.6 & 155.6 & 12.1$\pm$0.2 & 12.1$\pm$0.2 \\ 
Q1211p1030 & 0.3929 & 3 & 3 & 18.7 & 18.7 & 10.0$\pm$0.8 & 10.0$\pm$0.8 \\ 
Q1229m021 & 0.7691 & 3 & 3 & 256.7 & 256.7 & 12.4$\pm$0.2 & 12.4$\pm$0.2 \\ 
Q1229m021 & 0.8311 & 4 & 4 & 122.0 & 122.0 & 12.2$\pm$0.3 & 12.2$\pm$0.3 \\ 
Q2131m1207 & 0.4298 & 4 & 4 & 118.3 & 118.3 & 12.0$\pm$0.3 & 12.0$\pm$0.3 \\ 
Q0420m0127 & 0.6331 & 4 & 5 & 176.3 & 324.9 & 12.2$\pm$0.2 & 13.0$\pm$0.1 \\ 
Q1211p1030 & 0.8999 & 4 & 6 & 94.0 & 357.8 & 12.1$\pm$0.3 & 13.3$\pm$0.1 \\ 
Q0152p0023 & 0.4818 & 5 & 7 & 249.4 & 298.7 & 12.9$\pm$0.2 & 13.2$\pm$0.1 \\ 
Q0152m2001 & 0.383 & 6 & 7 & 315.6 & 350.9 & 12.9$\pm$0.1 & 13.1$\pm$0.1 \\ 
Q1515p0410 & 0.5592 & 6 & 7 & 170.0 & 188.5 & 12.6$\pm$0.2 & 12.7$\pm$0.2 \\ 
Q1130m1449 & 0.3127 & 14 & 16 & 183.7 & 255.7 & 12.4$\pm$0.2 & 12.8$\pm$0.2 \\ 
         \hline
    \end{tabular}
    
\end{table*}

\begin{figure}[h!]
    \centering
    \includegraphics[width=\linewidth]{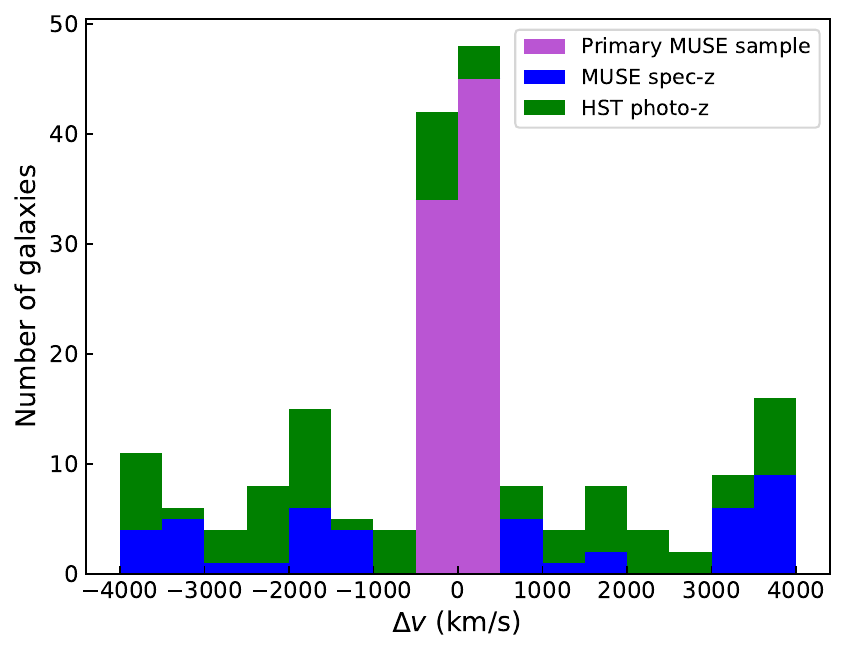}
    \caption{{The distribution of velocity separations between galaxies and absorbers across all systems. The colours distinguish the primary MUSE sources, MUSE sources with larger velocity separations, and HST sources.} }
    \label{fig:dv_hist}
\end{figure}

\begin{figure}[h!]
    \centering
    \begin{subfigure}[b]{0.49\textwidth}
        \centering
        \includegraphics[width=\linewidth]{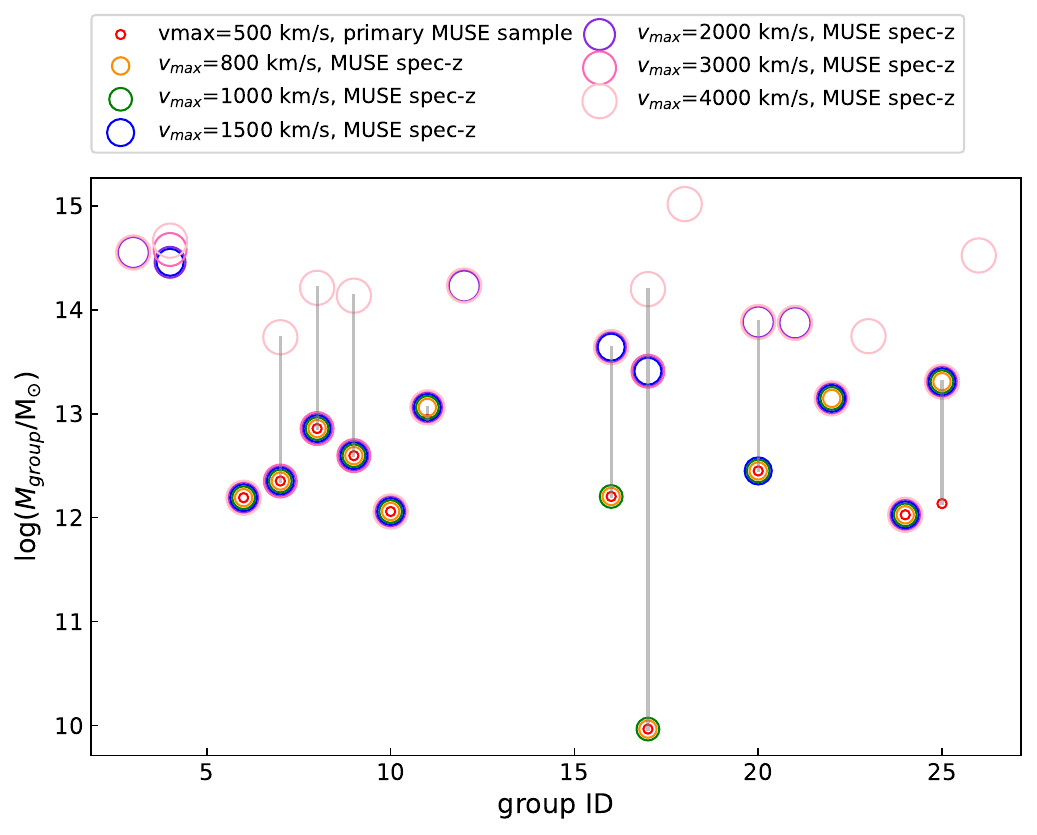}
        \caption{ }
        \label{fig:mgroup_specz}
    \end{subfigure}%
    
    \begin{subfigure}[b]{0.49\textwidth}
        \centering
        \includegraphics[width=\linewidth]{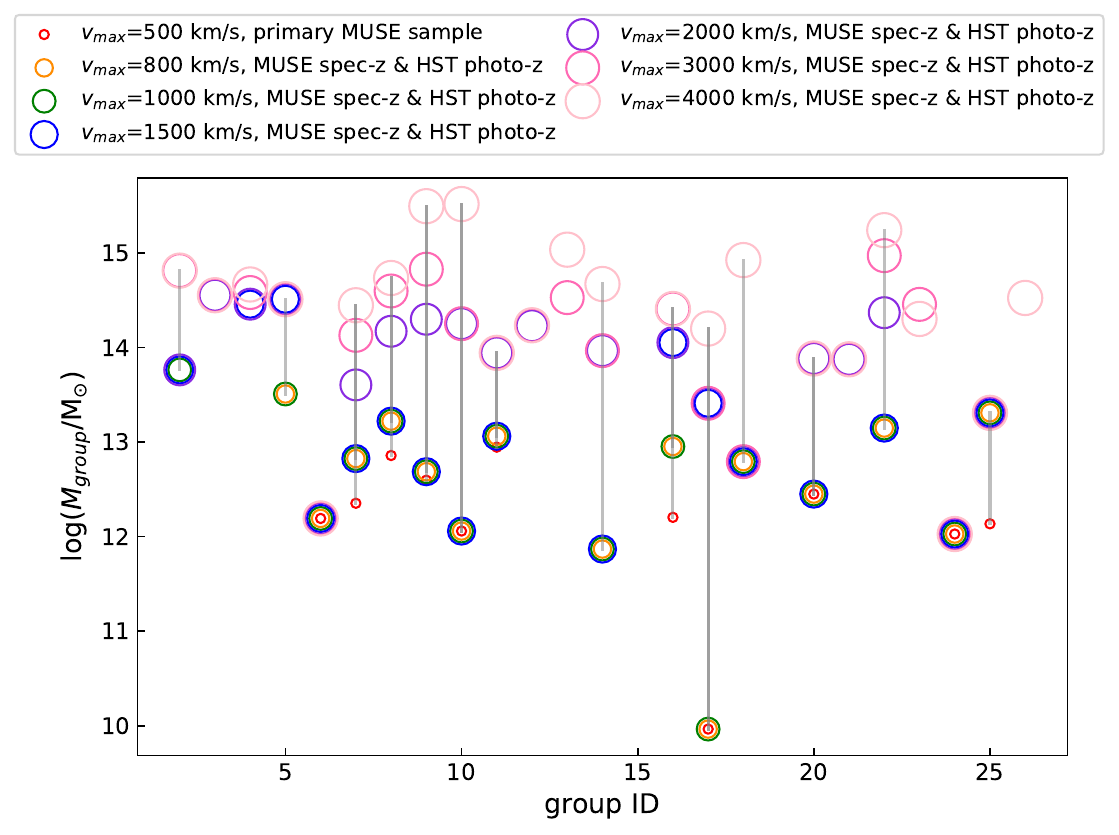}
        \caption{ }
        \label{fig:mgroup_both}
    \end{subfigure}
    
    \caption{{Derived group masses for different maximum velocity cuts, $v_{\rm max}$, used to define the groups. Panel (a) shows groups identified using only spectroscopic MUSE sources, while panel (b) includes both spectroscopic redshift MUSE and photometric redshift HST sources. Groups are shown in arbitrary order along the $x$-axis, with grey vertical lines connecting measurements for the same group. Circle colour and size indicate the value of $v_{\rm max}$ used. }}
    \label{fig:mgroup}
    \vspace{-5pt}
    \label{fig:mgroup}
\end{figure}

 {Figures \ref{fig:Q0138m0005_large} to \ref{fig:Q2131m1207_large} show the quasar fields and absorber systems. {Only the 16 of the 19 absorber fields with galaxies identified by MUSE at the absorber redshift are shown.} The background greyscale images are from HST. All galaxies {at the absorber redshift} identified with MUSE are marked with circles. {When multiple absorbers are present in the quasar spectrum, the circle colour indicates the absorber redshift with which each galaxy is associated.} Galaxies with CO detections from ALMA are additionally indicated by contours (for sources in the large programme) or diamond markers (for detections reported in previous studies). }

\begin{figure}[h!]
    \centering
    \includegraphics[width=0.77\linewidth]{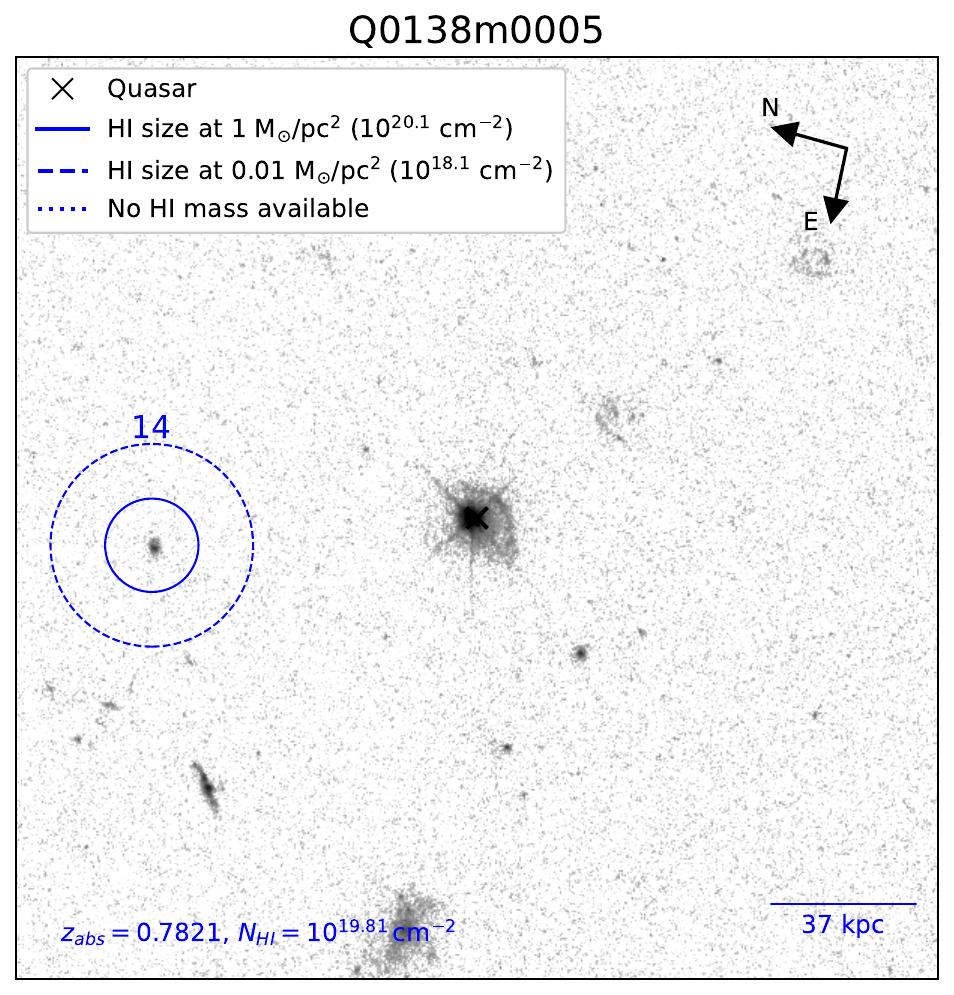}
    \caption{ {The quasar field and absorber-galaxy system. The background greyscale image is from HST. All galaxies at the absorber redshift identified with MUSE are marked with circles showing their inferred \HI\ size. {When multiple absorbers are present in the quasar spectrum, the circle colour indicates the absorber redshift with which each galaxy is associated.} When a galaxy has a CO detections from ALMA, it is indicated by contours (for sources in the large programme) or diamond markers (for detections reported in previous studies).} }
    \label{fig:Q0138m0005_large}
\end{figure}

\begin{figure}[h!]
    \centering
    \includegraphics[width=0.77\linewidth]{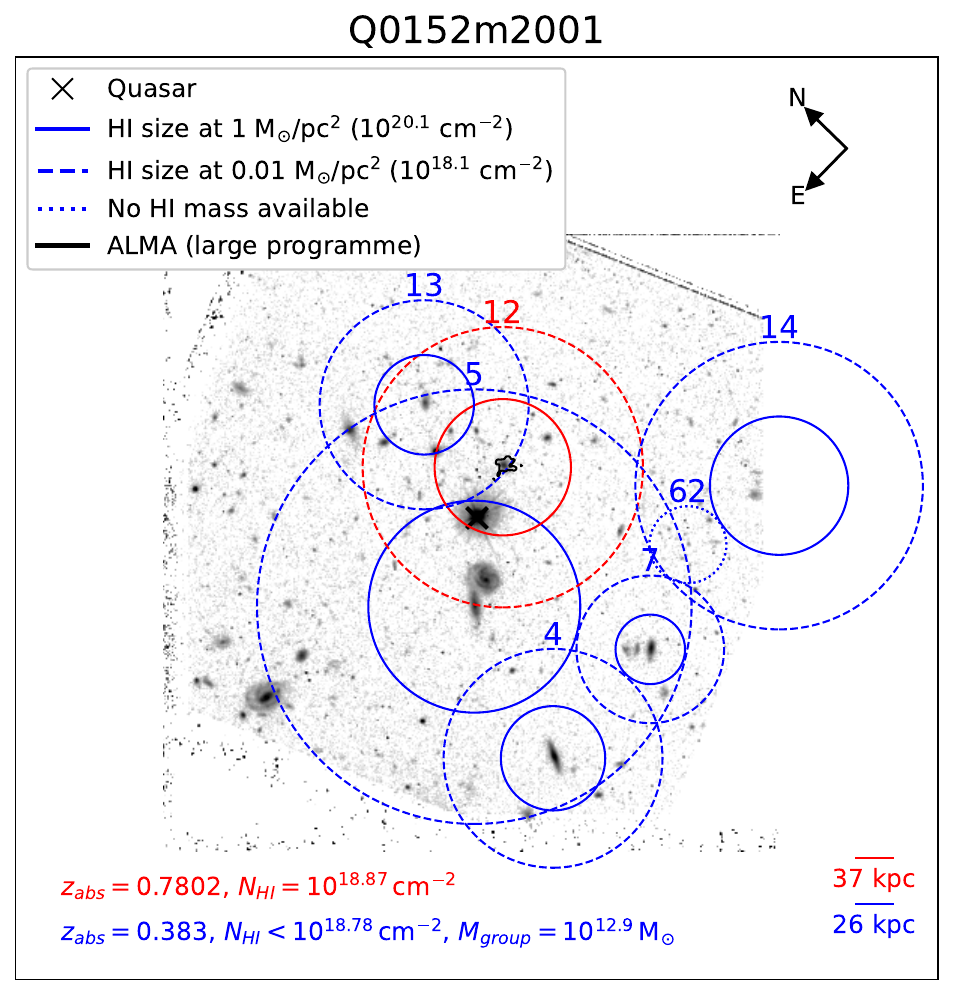}
    \caption{ {Same as Fig. \ref{fig:Q0138m0005_large}.} }
    \label{fig:Q0152m2001_large}
\end{figure}

\begin{figure}[h!]
    \centering
    \includegraphics[width=0.77\linewidth]{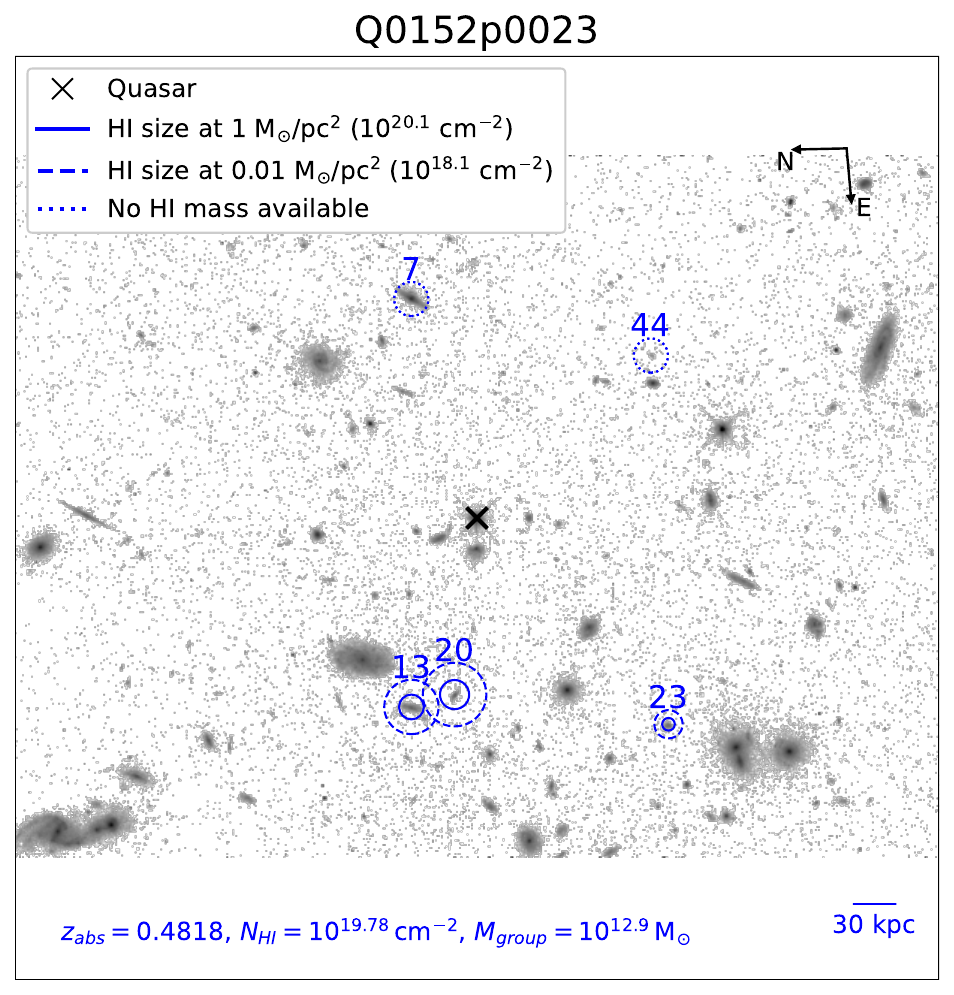}
    \caption{{Same as Fig. \ref{fig:Q0138m0005_large}.}  }
    \label{fig:Q0152p0023_large}
\end{figure}

\begin{figure}[h!]
    \centering
    \includegraphics[width=0.77\linewidth]{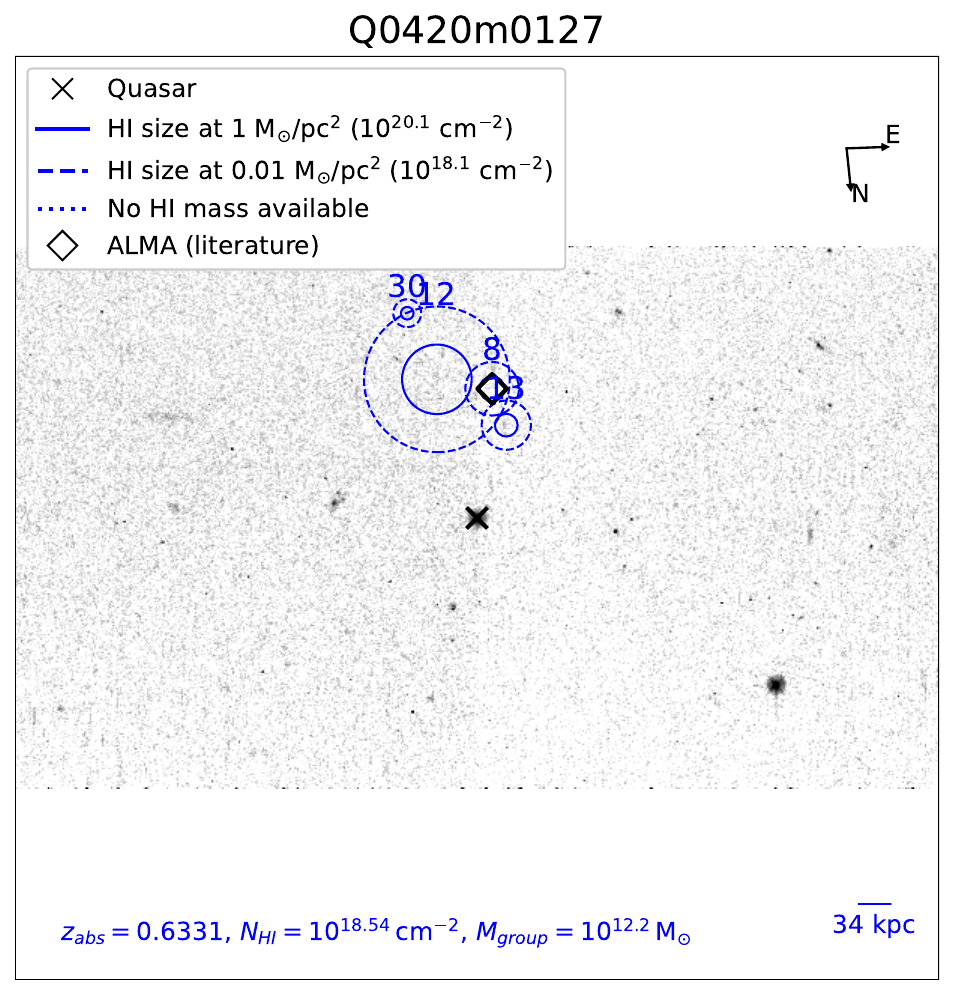}
    \caption{{Same as Fig. \ref{fig:Q0138m0005_large}.}  }
    \label{fig:Q0420m0127_large}
\end{figure}

\begin{figure}[h!]
    \centering
    \includegraphics[width=0.77\linewidth]{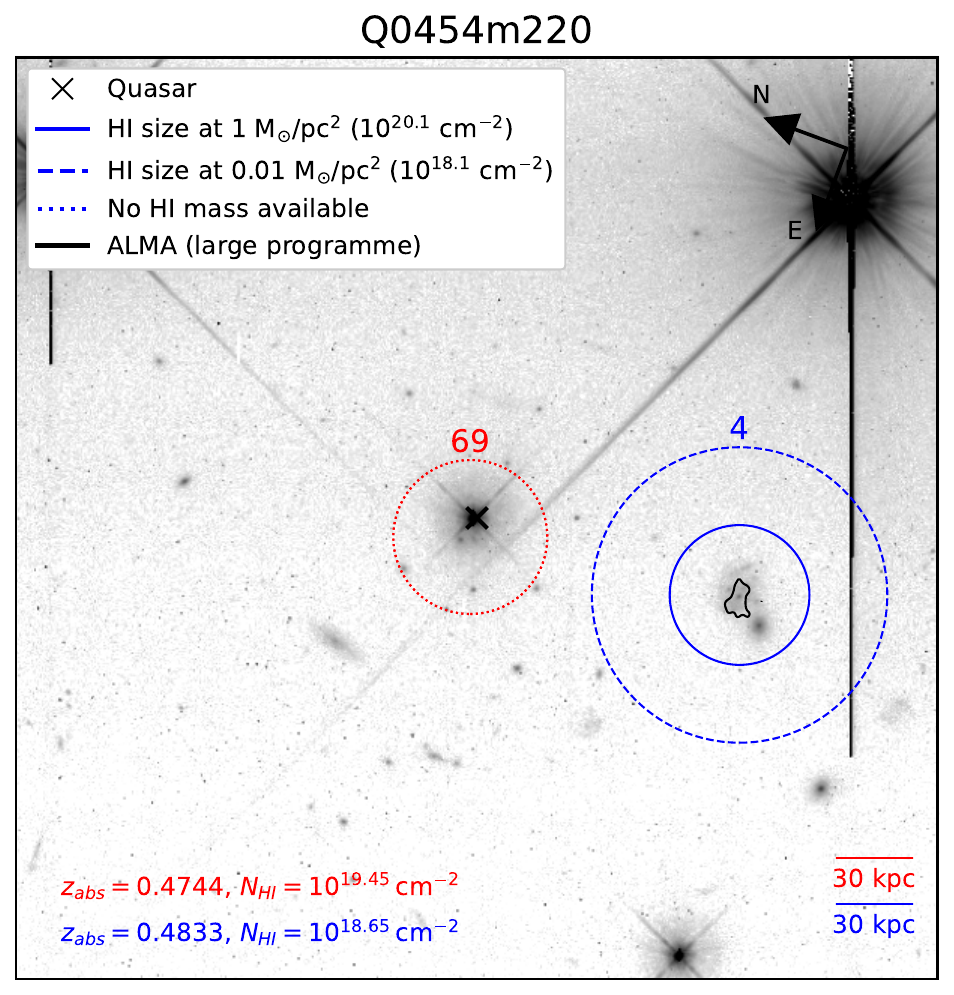}
    \caption{{Same as Fig. \ref{fig:Q0138m0005_large}.}  }
    \label{fig:Q0454m220_large}
\end{figure}

\begin{figure}[h!]
    \centering
    \includegraphics[width=0.77\linewidth]{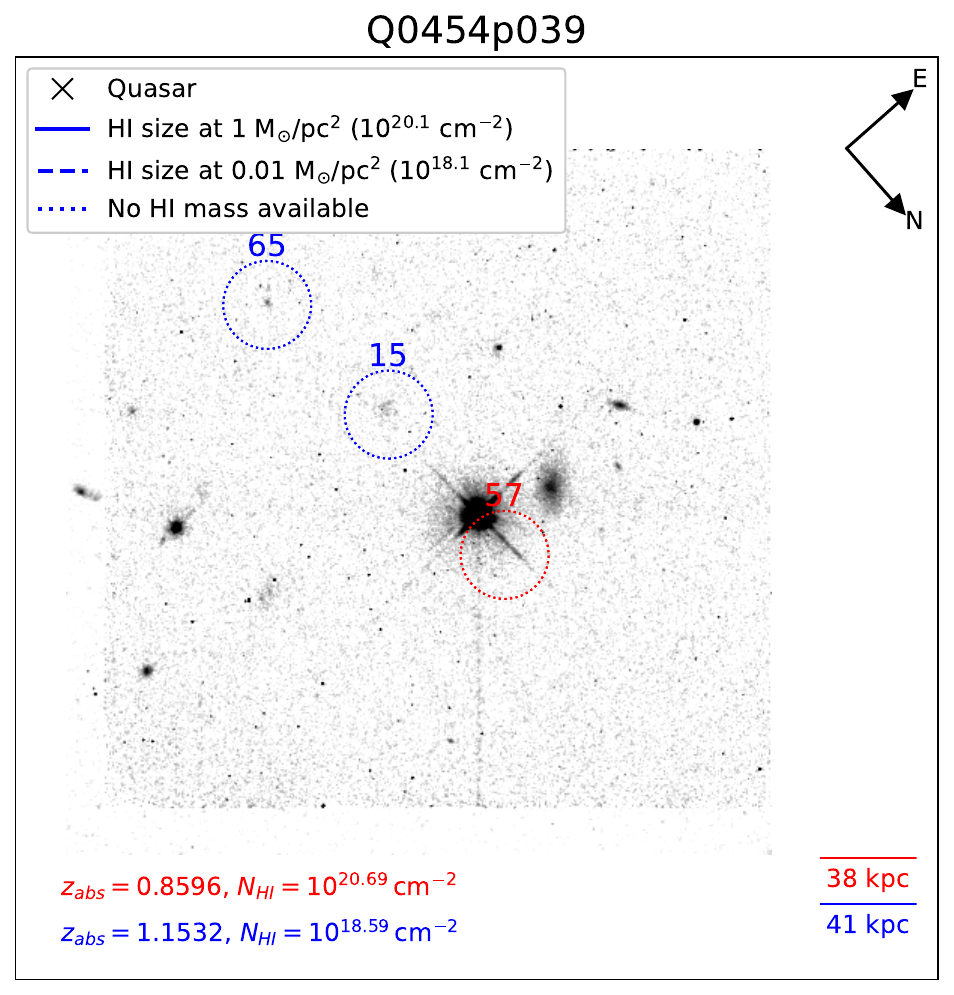}
    \caption{ {Same as Fig. \ref{fig:Q0138m0005_large}.} }
    \label{fig:Q0454p039_large}
\end{figure}

\begin{figure}[h!]
    \centering
    \includegraphics[width=0.77\linewidth]{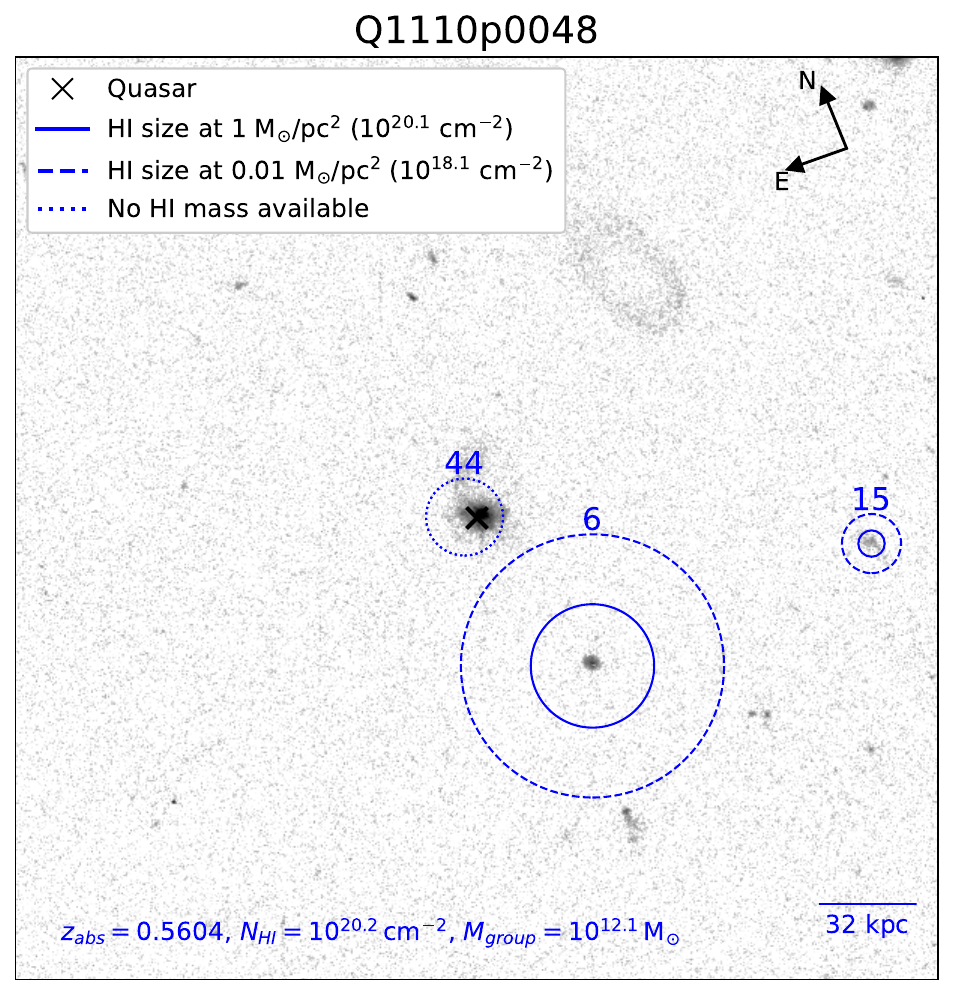}
    \caption{ {Same as Fig. \ref{fig:Q0138m0005_large}.} }
    \label{fig:Q1110p0048_large}
\end{figure}

\begin{figure}[h!]
    \centering
    \includegraphics[width=0.77\linewidth]{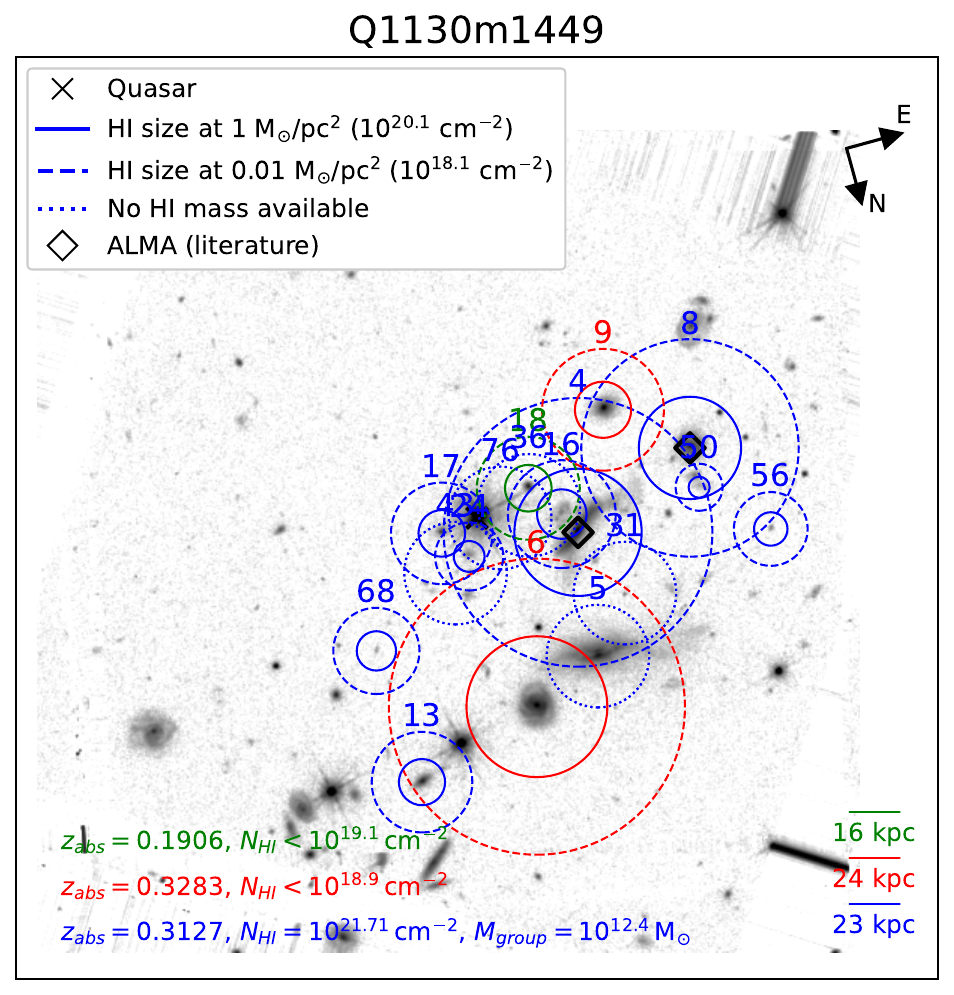}
    \caption{ {Same as Fig. \ref{fig:Q0138m0005_large}.} }
    \label{fig:Q1130m1449_large}
\end{figure}

\begin{figure}[h!]
    \centering
    \includegraphics[width=0.77\linewidth]{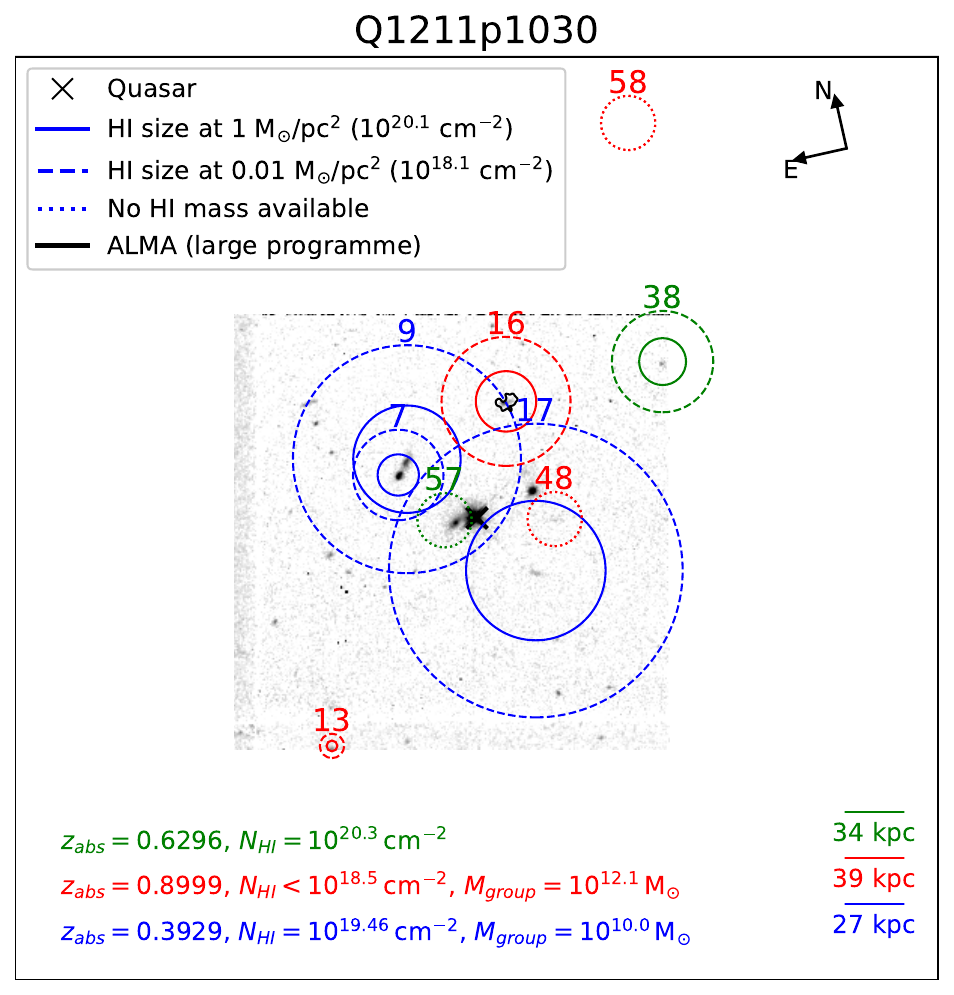}
    \caption{{Same as Fig. \ref{fig:Q0138m0005_large}.}  }
    \label{fig:Q1211p1030_large}
\end{figure}

\begin{figure}[h!]
    \centering
    \includegraphics[width=0.77\linewidth]{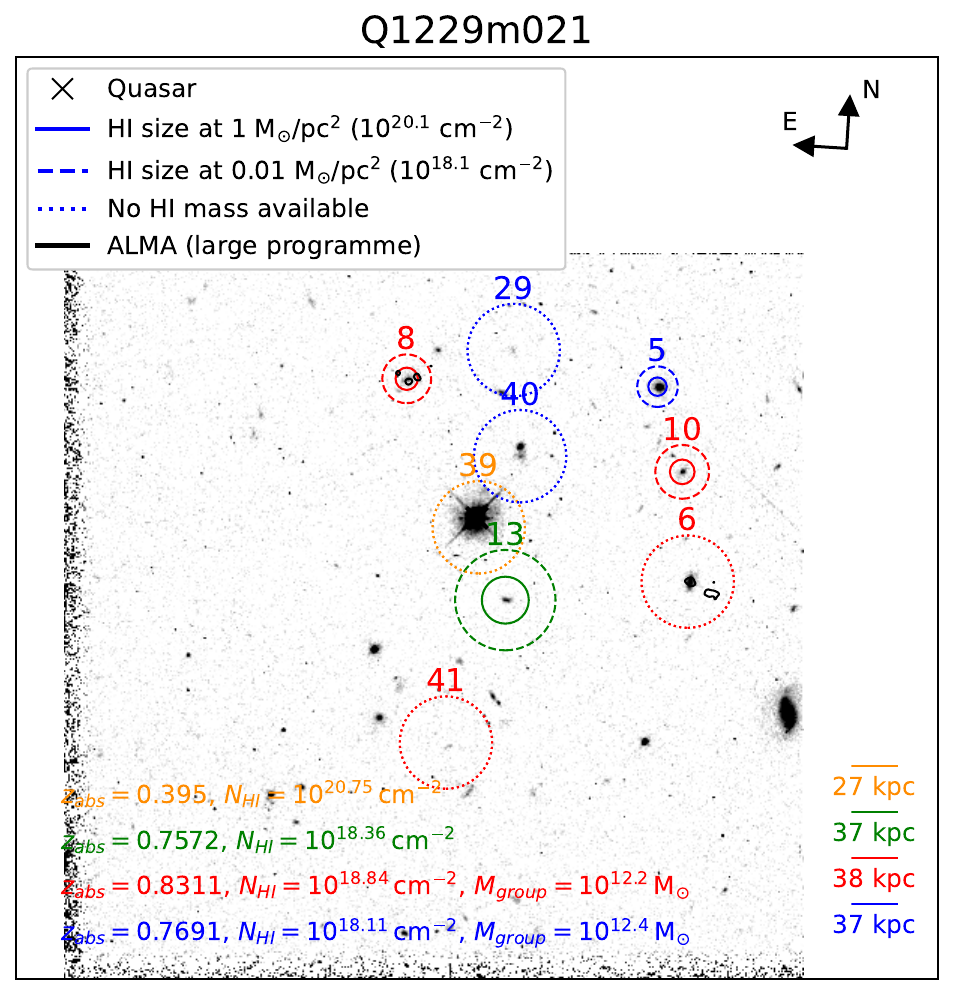}
    \caption{ {Same as Fig. \ref{fig:Q0138m0005_large}.} }
    \label{fig:Q1229m021_large}
\end{figure}

\begin{figure}[h!]
    \centering
    \includegraphics[width=0.77\linewidth]{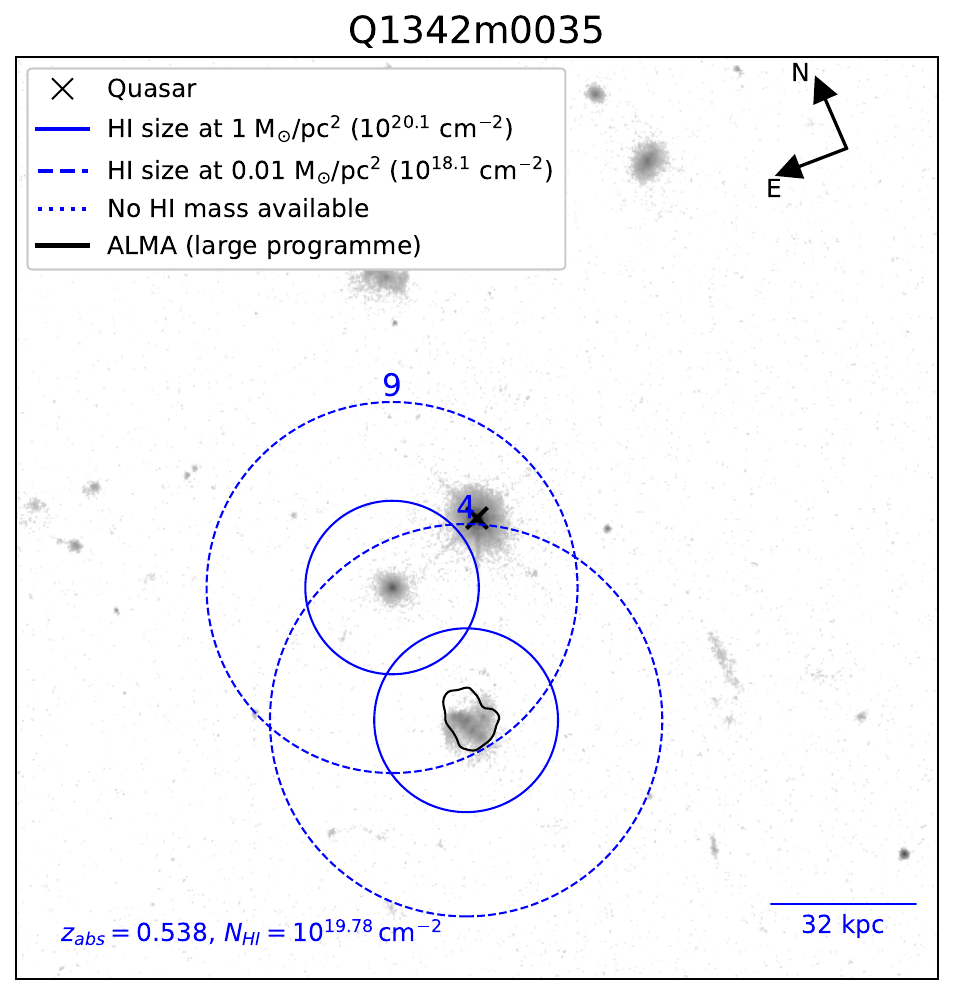}
    \caption{{Same as Fig. \ref{fig:Q0138m0005_large}.}  }
    \label{fig:Q1342m0035_large}
\end{figure}

\begin{figure}[h!]
    \centering
    \includegraphics[width=0.77\linewidth]{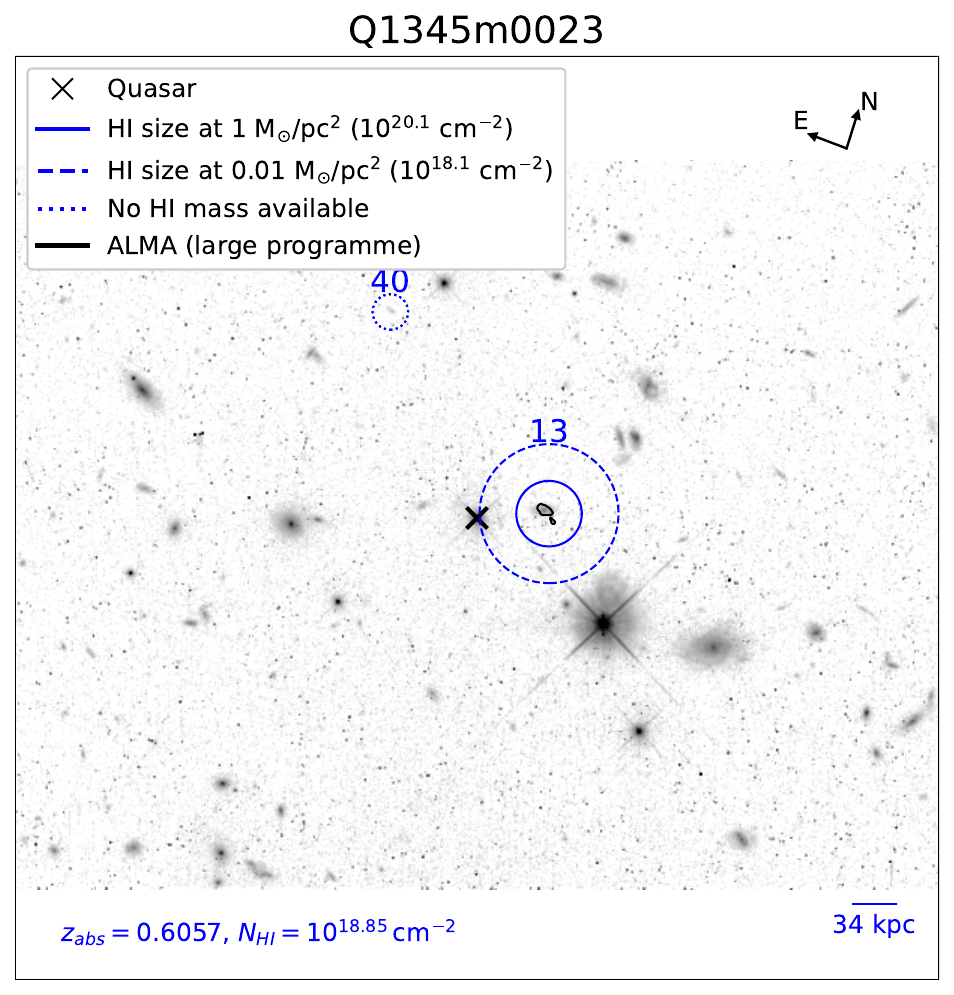}
    \caption{ {Same as Fig. \ref{fig:Q0138m0005_large}.} }
    \label{fig:Q1345m0023_large}
\end{figure}

\begin{figure}[h!]
    \centering
    \includegraphics[width=0.77\linewidth]{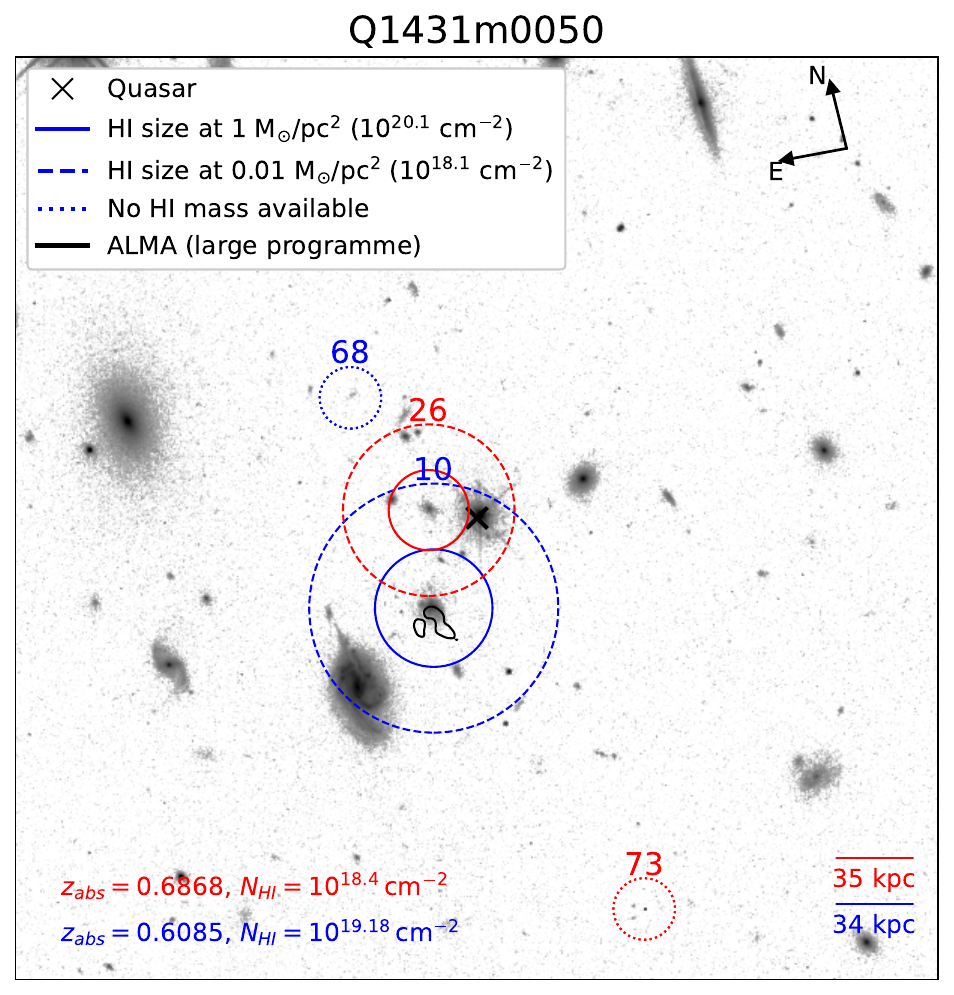}
    \caption{{Same as Fig. \ref{fig:Q0138m0005_large}.}  }
    \label{fig:Q1431m0050_large}
\end{figure}

\begin{figure}[h!]
    \centering
    \includegraphics[width=0.77\linewidth]{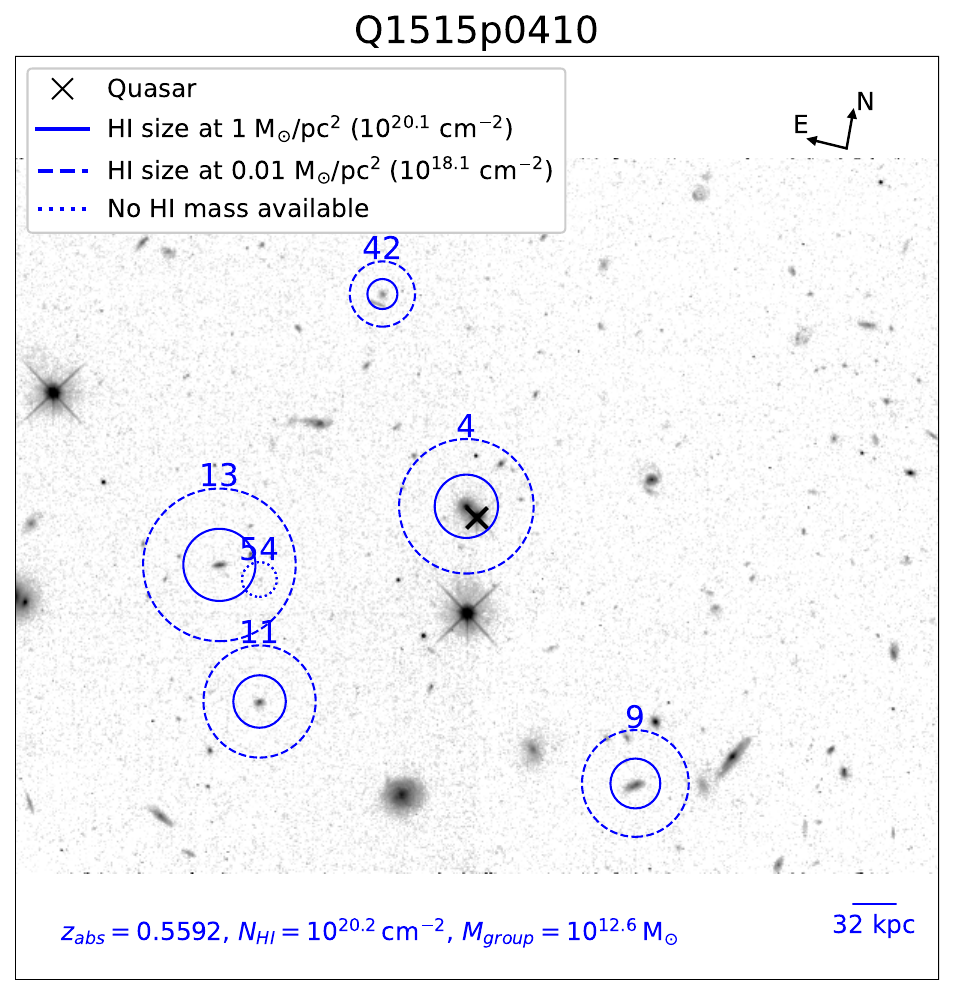}
    \caption{ {Same as Fig. \ref{fig:Q0138m0005_large}.} }
    \label{fig:Q1515p0410_large}
\end{figure}

\begin{figure}[h!]
    \centering
    \includegraphics[width=0.77\linewidth]{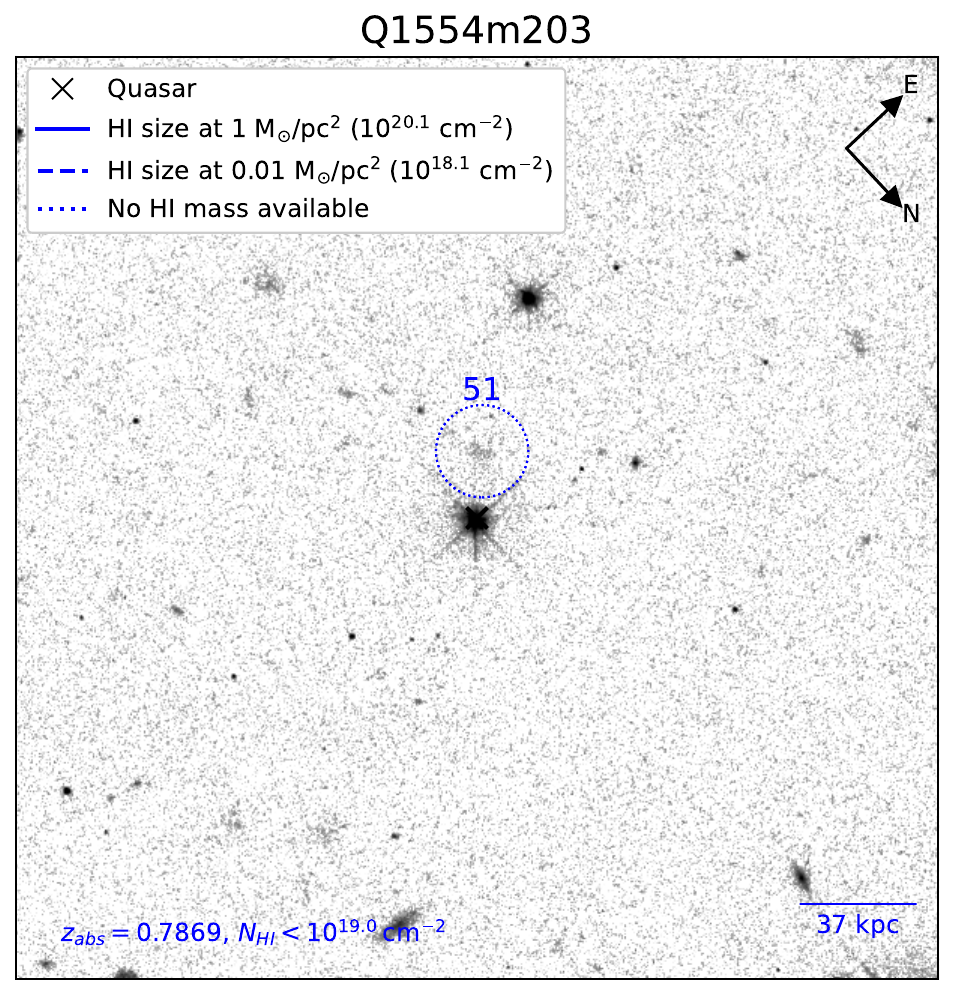}
    \caption{{Same as Fig. \ref{fig:Q0138m0005_large}.}  }
    \label{fig:Q1554m203_large}
\end{figure}

\begin{figure}[h!]
    \centering
    \includegraphics[width=0.77\linewidth]{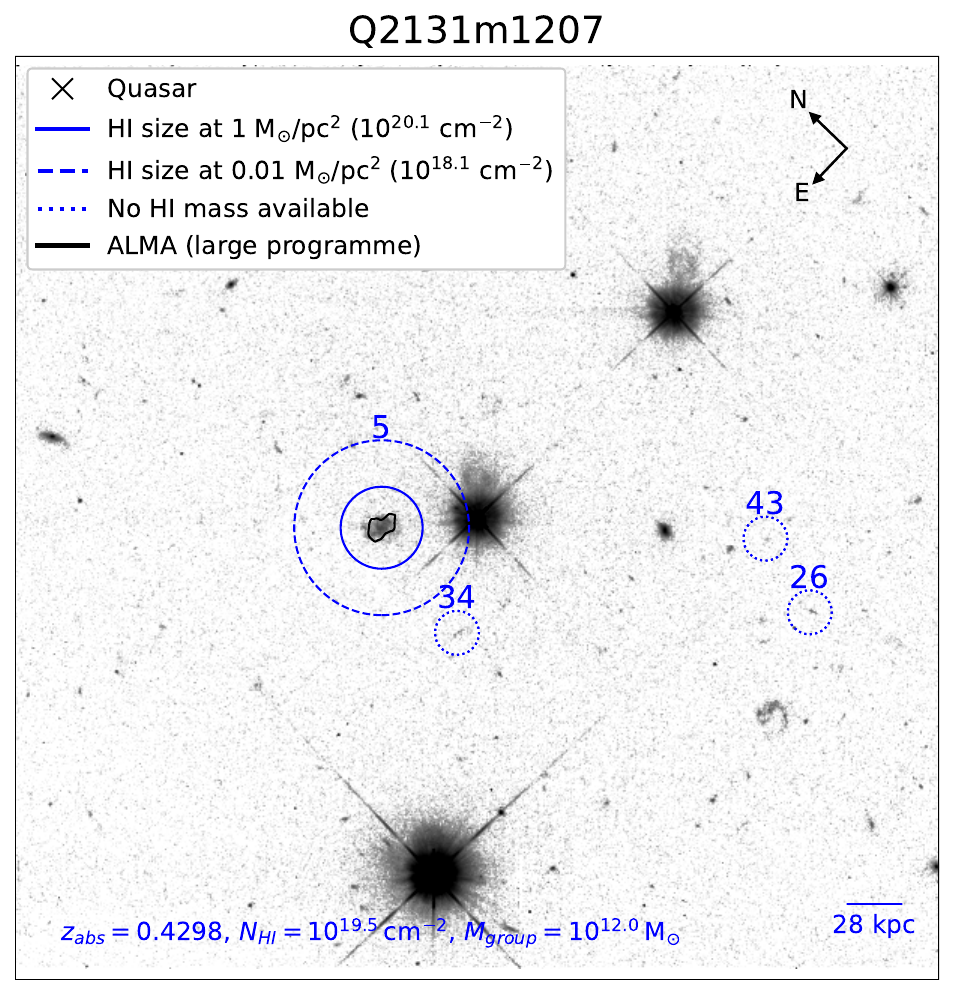}
    \caption{ {Same as Fig. \ref{fig:Q0138m0005_large}.} }
    \label{fig:Q2131m1207_large}
\end{figure}

\section{Gas fraction}
\label{appen:gf}

 An alternative representation of the gas fraction (Sect. \ref{sec:gas_frac}) is $f_{gas}=M_{\rm gas}/(M_{\rm gas}+M_{*})$, shown in Fig. \ref{fig:fg_appen}. Using the scaling relations from \cite{Tacconi:20}, we recreate the plot presented in \cite{ForsterScheiber:20} showing the MUSE-ALMA Haloes galaxies with respect to the redshift evolution of the molecular gas fraction. Just as we see with the molecular gas to stellar mass ratio in Fig. \ref{fig:fgas}, the MUSE-ALMA Haloes galaxies have lower molecular gas content for their stellar mass than is expected for $z\sim0.5$ galaxies, more consistent with the $z\sim0$ relation. However, this comparison assumes $M_{\rm gas} = M_{\rm H2}$. If the \HI\ component is included, the galaxies appear significantly more gas rich, shifting them closer to the $z\sim1$ relation. This further highlights that \HI\ can constitute a substantial fraction of the cold gas reservoir in these systems.

 \begin{figure}[h!]
    \centering
    \begin{subfigure}[b]{0.5\textwidth}
        \centering
        \includegraphics[width=\linewidth]{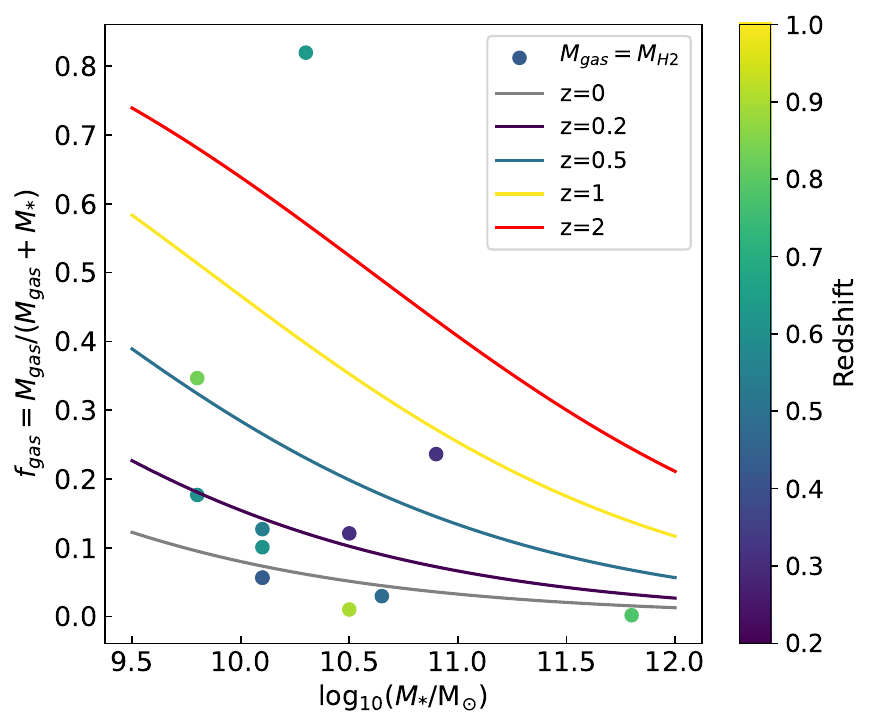}
        \caption{ }
        \label{fig:fb1}
    \end{subfigure}%

    \begin{subfigure}[b]{0.5\textwidth}
        \centering
        \includegraphics[width=\linewidth]{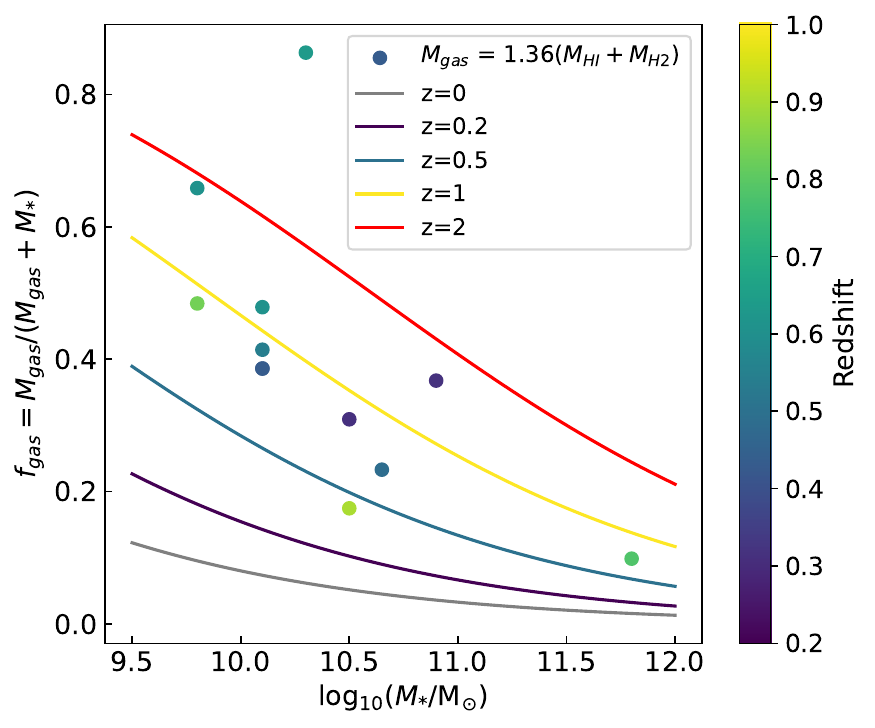}
        \caption{ }
        \label{fig:fb2}
    \end{subfigure}
    
    \caption{{Gas fraction as a function of stellar mass, illustrating the expected redshift evolution and the effect of including both molecular and atomic gas. The MUSE-ALMA Haloes galaxies are shown as points coloured by redshift, with panel (a) displaying the molecular gas fraction, $M_{\rm gas}=M_{\rm H2}$, and panel (b) displaying the total gas fraction, $M_{\rm gas}=M_{\rm HI}+M_{\rm H2}$. In both panels, the coloured curves show the molecular gas fraction scaling relations as a function of stellar mass at different redshifts from \citet{Tacconi:20}, as presented by \citet{ForsterScheiber:20}. }  }
    \label{fig:fg_appen}
\end{figure}

\end{appendix}

\end{document}